\documentclass[conference,letterpaper]{IEEEtran}
\IEEEoverridecommandlockouts
\usepackage{epsfig}
\usepackage{times}
\usepackage{float}
\usepackage{stfloats}
\usepackage{afterpage}

\usepackage{amstext,cite}
\usepackage{amssymb,bm}
\usepackage{latexsym}
\usepackage{color}
\usepackage{graphicx}

\usepackage{amsthm}
\usepackage{graphicx}
\usepackage[center]{caption}
\usepackage{pstricks}
\usepackage{caption}
\usepackage{subcaption}
\usepackage{booktabs}
\usepackage{multicol}
\usepackage{lipsum}
\usepackage[T1]{fontenc}
\usepackage{hyperref}
 \usepackage{aecompl}
 \usepackage{mathrsfs}
\usepackage{arydshln}  

\long\def\comment#1{}

\newfont{\bbb}{msbm10 scaled 700}

\newfont{\bb}{msbm10 scaled 1100}

\newcommand{\Ac}{{\cal A}}

\newcommand{\Gc}{{\cal G}}

\newcommand{\Pc}{{\cal P}}
\newcommand{\Qc}{{\cal Q}}

\newcommand{\Sc}{{\cal S}}

\newcommand{\Wc}{{\cal W}}

\newcommand{\Zc}{{\cal Z}}

\newcommand{\qsf}{{\sf q}}

\newcommand{\xsf}{{\sf x}}

\newcommand{\Ksf}{{\sf K}}
\newcommand{\Lsf}{{\sf L}}

\newcommand{\Rsf}{{\sf R}}

\newcommand{\Tsf}{{\sf T}}

\newcommand{\be}{\begin{equation}}
\newcommand{\ee}{\end{equation}}
\newcommand{\bea}{\begin{eqnarray}}
\newcommand{\eea}{\end{eqnarray}}
\newtheorem{thm}{Theorem}
\newtheorem{cor}{Corollary}

\newtheorem{rem}{Remark}
\newtheorem{example}{Example}

\usepackage[utf8]{inputenc} 
\usepackage[T1]{fontenc}
\usepackage{url}
\usepackage{ifthen}
\usepackage{cite}
\usepackage[cmex10]{amsmath} 

\usepackage{amsmath}
\usepackage{tikz}
\usetikzlibrary{calc}

\usepackage[nodisplayskipstretch]{setspace} 
\newcommand{\tikzmark}[1]{\tikz[overlay,remember picture] \node (#1) {};}
\newcommand{\DrawboxF}[4][]{%
    \tikz[overlay,remember picture]{%
        \coordinate (TopLeft)     at ($(#2)+(-0.2em,1.5em)$); 
        \coordinate (BottomRight) at ($(#3)+(0.2em,-2.5em)$); %
        \path (TopLeft); \pgfgetlastxy{\XCoord}{\IgnoreCoord};
        \path (BottomRight); \pgfgetlastxy{\IgnoreCoord}{\YCoord};
        \coordinate (LabelPoint) at ($(\XCoord,\YCoord)!0.5!(BottomRight)$);
        \draw [#1, dashed] (TopLeft) rectangle (BottomRight); 
       \node [#1, fill=none, fill opacity=1] at ([xshift=-2em, yshift=-6pt]BottomRight.south) {#4};
    }
}

\newcommand{\DrawboxV}[4][]{%
    \tikz[overlay,remember picture]{%
        \coordinate (TopLeft)     at ($(#2)+(-0.2em,1.5em)$);
        \coordinate (BottomRight) at ($(#3)+(0.2em,-3.5em)$); 
        \path (TopLeft); \pgfgetlastxy{\XCoord}{\IgnoreCoord};
        \path (BottomRight); \pgfgetlastxy{\IgnoreCoord}{\YCoord};
        \coordinate (LabelPoint) at ($(\XCoord,\YCoord)!0.5!(BottomRight)$);
        \draw [#1, dashed] (TopLeft) rectangle (BottomRight);
       \node [#1, fill=none, fill opacity=1] at ([xshift=-2em, yshift=-10pt]BottomRight.south) {#4};
    }
}

\begin{document}
\title{On the Optimal Source Key Size of Secure Gradient Coding} 


\author{%
  \IEEEauthorblockN{Yang Zhou, Wenbo Huang, Kai Wan, Robert Caiming Qiu}
  \IEEEauthorblockA{The School of Electronic Information and Communications,\\
                    Huangzhong University of Science and Technology,\\
                    430074 Wuhan, China\\
                    Email: \{hust\_zhou, eric\_huang, kai\_wan, caiming\}@hust.edu.cn}
  \and
  \IEEEauthorblockN{Claude E.~Shannon and David Slepian}
  \IEEEauthorblockA{Bell Telephone Laboratories, Inc.\\ 
                    Murray Hill, NJ, USA\\
                    Email: \{csh, dsl\}@bell-labs.com}
}

\author{
\IEEEauthorblockN{%
Yang Zhou\IEEEauthorrefmark{1},
Wenbo Huang\IEEEauthorrefmark{1}\textsuperscript{\dag}\thanks{\dag~Corresponding author: Wenbo Huang (eric\_huang@hust.edu.cn)},
Kai Wan\IEEEauthorrefmark{1},
Robert Caiming Qiu\IEEEauthorrefmark{1}
}
\IEEEauthorblockA{\IEEEauthorrefmark{1}Huazhong University of Science and Technology, 430074  Wuhan, China, \\ \{hust\_zhou, eric\_huang, kai\_wan, caiming\}@hust.edu.cn}
}

\maketitle


\begin{abstract}
Gradient coding enables a user node to efficiently aggregate gradients computed by server nodes from local datasets, achieving low communication costs while ensuring resilience against straggling servers. This paper considers the secure gradient coding problem, where a user aims to compute the sum of the gradients from ${\sf K}$ datasets with the assistance of ${\sf N}$ distributed servers. The user is required to recover the sum of gradients from the transmissions of any $\mathsf{N}_\mathrm{r}$ servers, with each dataset assigned to $\mathsf{N}-\mathsf{N}_\mathrm{r}+\mathsf{m}$ servers. The security constraint guarantees that even if the user receives transmissions from all servers, no additional information about the datasets can be obtained beyond the sum of gradients. It has been shown in the literature that the security constraint does not increase the optimal communication cost of the gradient coding problem, provided that enough source keys are shared among the servers. However, the minimum required source key size to ensure security while achieving the optimal communication cost has been studied only for the case $\mathsf{m}=1$. In this paper, we focus on the more general case $\mathsf{m}\geq 1$ and aim to characterize the minimum required source key size for this purpose. A new information-theoretic converse bound on the source key size and a novel achievable scheme with smartly designed assignments are proposed. Our proposed scheme outperforms the optimal scheme based on the widely used cyclic data assignment and coincides with the converse bound under specific system parameters.
\end{abstract}

\section{Introduction}

Secure aggregation is a well-established topic in distributed computation that addresses the multiparty computation problem in which the goal is to compute a sum over multiple parties without revealing individual updates, not even to the aggregator~\cite{Bonawitz_Secure_Aggregation}. This primitive has already seen widespread deployment in privacy‑sensitive applications, ranging from on‑device next‑word prediction in mobile keyboards~\cite{Bonawitz_Secure_Aggregation} to cross‑institutional medical‑model training~\cite{sho2015}. 
The concept of secret sharing was first introduced in \cite{shamir1979share}, where a secret is divided into multiple shares so that only participants meeting a predefined threshold can reconstruct the secret. 
The complexity of verifiable secret sharing and its extension to multiparty computation was further investigated in \cite{pedersen1991non, cramer2000complexity, ChaumMultiparty1988,yue2016healthcare}.

In this paper, we consider the secure aggregation in gradient coding, called secure gradient coding. 
The gradient coding problem was first proposed in~\cite{pmlr-v70-tandon17a}. 
In a general gradient coding model, a user wants to compute a sum of gradients on   ${\sf K}$ datasets by ${\sf N}$ servers. During the data assignment phase, each dataset is assigned to $\mathsf{M} = \mathsf{N} - \mathsf{N}_\mathrm{r} + \mathsf{m}$  servers.  During the computing phase, each server first computes gradients of the assigned datasets and then sends a coded message on these gradients to the user. During the decoding phase, the user should recover the computation task result using the transmissions from any ${\sf N_r}$ servers. The objective is to minimize the number of transmissions (i.e., communication cost). 
In~\cite{ye2018communication}, the optimal communication cost under linear encoding was characterized. 
Furthermore, the authors in~\cite{Heterogeneous_Gradient_Coding} consider heterogeneous (and arbitrarily fixed) data assignment, where each dataset may be assigned to a different number of servers,  and 
characterizes the optimal communication cost under linear encoding. 

The secure aggregation model for gradient coding was first proposed in~\cite{wan2022secure}.\footnote{\label{foot:linearly separable}The secure distributed linearly separable computation was formulated in~\cite{wan2022secure}. When the user only requests one linear combination of the gradients (or input vectors),  the problem reduces to secure gradient coding.} 
In this model, a key distribution server assigns a shared key to all servers to ensure security.
It was proven in~\cite{wan2022secure} that the security constraint will not increase the optimal
communication cost of the gradient coding problem, if enough
source keys are shared among the servers. Then 
for the case  $\mathsf{m}=1$ (i.e., when $\mathsf{M}$ is minimum to tolerate $\mathsf{N}-\mathsf{N}_\mathrm{r}$ stragglers), an information-theoretic lower bound on the source key size and several secure distributed coded computation schemes for specific data distribution patterns (e.g., cyclic, repetitive, and mixed) were proposed in~\cite{wan2022secure},  while achieving the optimal communication cost with different requirement on the source key size.

Recently,  secure aggregation for federated learning has also gained significant attention \cite{fereidooni2021safe,choi2020communication, lightsecagg,jahani2022swiftagg,zhao2022information}. 
In secure aggregation for federated learning, each distributed computing node holds its local data to compute the gradient; thus, secure aggregation against stragglers (or user dropouts) requires two-round transmissions. In contrast, in the gradient coding problem, an additional data assignment phase for servers exists. Because of the redundancy in data assignment, one-round transmission is enough for secure aggregation.

\paragraph*{Main Contributions} 
In this paper, we aim to characterize the minimum source key size for the case $\mathsf{m}>1$ while achieving the optimal communication cost under linear encoding.  Our main contributions are as follows.
\begin{itemize}
\item For a given data assignment, we provide an information-theoretic converse bound on the source key size, while achieving the optimal communication cost under linear encoding.
\item To reduce the source key size, we introduce a secure gradient coding scheme with a new data assignment strategy. This scheme outperforms the widely used cyclic assignment scheme and can achieve the converse bound under some system parameters.

\end{itemize}

\paragraph*{Notation Convention}
Calligraphic symbols denote sets,  
bold lower-case letters denote vectors, bold upper-case letters denote matrices, and sans-serif symbols denote system parameters.
We use $|\cdot|$ to represent the cardinality of a set or the length of a vector;
$[a:b]:=\left\{ a,a+1,\ldots,b\right\}$; 
  $[n] := [1:n]$;
$\mathbb{F}_{\qsf}$ represents a  finite field with order $\qsf$;         
$\mathbf{M}^{\text{T}}$  and $\mathbf{M}^{-1}$ represent the transpose  and the inverse of matrix $\mathbf{M}$, respectively;
 the matrix $[a;b]$ is written in a Matlab form, representing $[a,b]^{\text{T}}$;
$(\mathbf{M})_{m \times n}$ represents the dimension of matrix $\mathbf{M}$ is $m \times n$;
 $\text{Mod} (b,a)$ represents the modulo operation on $b$ with  integer divisor $a$ and in this paper we let $\text{Mod}(b,a)\in \{1,\ldots,a \}$ (i.e., we let $ \text{Mod}(b,a)=a$ if $a$ divides $b$).

\section{System Model}
\label{sec:system}
This section describes the $(\Ksf,\mathsf{N},\mathsf{N}_\mathrm{r}, \mathsf{m})$ secure gradient coding problem. 
We consider a framework in which a user, assisted by $\mathsf{N}$ servers, aims to compute the sum of gradients derived from $\Ksf$ independent datasets $D_1,\ldots, D_{\Ksf}$, where each gradient represented as a string of $\Lsf$ symbols over a sufficiently large alphabet. 
 Compared to distributed computing without considering secure aggregation, there is another trusted server to distribute keys to the computing servers. It generates a set of  random variables $\Qc = \{Q_{1}, \ldots, Q_{\mathsf{N}}\}$ on $\mathbb{F}_{\qsf}$ independent of the datasets, 
\begin{align}
I(Q_1, \ldots, Q_{\mathsf{N}} ; D_1, \ldots, D_{\Ksf} ) = 0. \label{eq:independent key}
\end{align}
The source key size \( \eta \) measures the total amount of randomness among all keys, i.e.,
\begin{align}
\eta = H( Q_1, \ldots, Q_{\mathsf{N}} )/\Lsf. \label{eq:def of eta}
\end{align}

The secure gradient coding framework comprises three phases: {\it data assignment, computing, and decoding}. 

{\it Data assignment phase.}
 We define ${\sf N_r}$ as the number of minimum available servers, and the identity of the surviving servers is not known until the decoding phase. 
The data center assigns each dataset to $\mathsf{M} =\mathsf{N}-\mathsf{N}_\mathrm{r}+\mathsf{m}$ servers, where $\mathsf{m}$ is the computation cost factor (which is also called communication reduction factor in~\cite{ye2018communication}). The index set of datasets assigned to server $n \in [{\sf N}]$ is denoted as $\Zc_n \in [\Ksf]$.  In addition, the trusted key server allocates the key   $\Qc_i$ to each server $i\in [\mathsf{N}]$.


{\it Computing phase.}
Each server $n \in[\mathsf{N}]$ first computes the message $g_k = \nabla (D_k)$ for each $k \in \Zc_n$. Then it sends the coded message to the user, 
\begin{align}
 X_n = \psi_n \left( \{g_k:  k \in \Zc_n\},  Q_n \right),
 \end{align}
  where  $\psi_n$  is a  function of messages $\{g_k: k\in \Zc_n\}$, 
\begin{align} 
\psi_n &:  \mathbb{F}_{\qsf}^{ |\Zc_n| \Lsf} \times   | \Qc |  \to \mathbb{F}_{\qsf}^{ \Tsf_n },  
\label{eq: encoding function def}
\end{align}
and $\Tsf_n$ represents the length of $ X_n $. 
$\mathsf{m}$ is the computation cost factor because in a distributed system, the complexity of computing the linear combinations of the gradients is much lower than computing the gradients. 

{\it Decoding phase.}
The user should be able to use coded messages from any $\mathsf{N}_\mathrm{r}$  servers to decode the computation tasks. 
Thus, for each subset of servers $\Ac \subseteq [\mathsf{N}]$ where $|\Ac| = \mathsf{N}_\mathrm{r}$, by defining
$
    {X_{\Ac}}: = ( {X_n}:n \in \Ac), 
$
it should hold that
\begin{align}
    H(g_1+\ldots+g_{\Ksf} | X_{\Ac}) = 0.
\end{align}

We define 
\begin{equation}
    \Rsf: = \mathop {\max }\limits_{\Ac \subseteq [\mathsf{N}]:|\Ac| = {\mathsf{N}_\mathrm{r}}} \frac{{\sum\nolimits_{n \in \Ac} {{\Tsf_n}} }}{\Lsf}
\end{equation} 
as the communication cost, which represents the maximum normalized number of symbols received from any $\mathsf{N}_\mathrm{r}$ servers to recover the computational task.

The secure aggregation protocol imposes that  
\begin{align}
    I\left(g_{1},\ldots, g_{\Ksf};   X_{[\mathsf{N}]} | g_1+\ldots+g_{\Ksf}  \right) = 0, \label{eq:security}
\end{align}
where $X_{[\mathsf{N}]}$ presents all messages from $\mathsf{N}$ servers which may be received by the user. This equation ensures that the user cannot obtain any information about the datasets except for the computation task.

If the source key size is large enough, the optimal communication cost under linear encoding (i.e., $ \psi_n$ is a linear function), can be characterized by directly combining~\cite{wan2022secure,ye2018communication}.\footnote{\label{foot:direct optimal}Without security, the authors in \cite{ye2018communication} characterized the optimal communication cost under linear encoding is $\mathsf{N}_\mathrm{r}/\mathsf{m}$. In addition,~\cite[Theorem 1]{wan2022secure} showed that for any $(\Ksf, \mathsf{N}, \mathsf{N}_\mathrm{r}, \mathsf{m})$ non-secure gradient coding problem, the coding scheme can be made secure without increasing the communication cost.} 
\begin{thm}[{\hspace{-0.04em}\cite{wan2022secure,ye2018communication}}]
\label{thm:communication cost}
For the  $(\Ksf,\mathsf{N},\mathsf{N}_\mathrm{r}, \mathsf{m})$ secure gradient coding problem, the optimal communication cost under linear encoding is 
$ \frac{\mathsf{N}_\mathrm{r}}{\mathsf{m}}$.
\end{thm}

Our main objective in this paper is to minimize the source key size $\eta$ while maintaining the optimal communication cost under linear encoding for the case $ \mathsf{m} \geq 1$.


\section{Main Results}
\label{sec:main}
We first introduce a novel converse bound on $ \eta $ for a fixed assignment, whose proof can be found in Appendix~\ref{sec:bound proof}.
\begin{thm}
\label{thm:bound of η}
For the $(\Ksf,\mathsf{N},\mathsf{N}_\mathrm{r}, \mathsf{m})$ secure gradient coding problem, for a fixed assignment \(\mathbf{Z} = (\Zc_1, \ldots, \Zc_{\mathsf{N}})\), if there exists an ordered set of servers in \([\mathsf{N}]\), denoted by \(\mathbf{s} = (s_1, \ldots, s_{|\mathbf{s}|})\), where each server contains at least one dataset that appears in the datasets of its preceding servers at most \(\mathsf{m} - 1\) times, this can be expressed as:
\begin{align}
\forall i \in [|\mathbf{s}|], \exists x \in \Zc_{s_i} \text{ such that } \sum_{j=1}^{i-1} \delta(x \in \Zc_{s_j}) \leq \mathsf{m} - 1, \label{eq:vector constraint}
\end{align}
where \(\delta(x \in \Zc_{s_j})\) equals \(1\) if \(x \in \Zc_{s_j}\), and \(0\) otherwise.

Under the symmetric transmission and linear coding, it must hold that
\begin{align}
\eta \geq \frac{|\mathbf{s}|}{\mathsf{m}}-1. \label{eq:converse lemma}
\end{align}
\end{thm}
Theorem~\ref{thm:bound of η} indicates that minimizing source key size $\eta$ is a combinatorial problem that depends on the data assignment. 
Further, we can establish a min-max problem to characterize the minimum $\eta^{\star}$ as follows.
\begin{cor}
    For the $(\Ksf,\mathsf{N},\mathsf{N}_\mathrm{r}, \mathsf{m})$ secure gradient coding problem, we have
    \begin{align}
\eta^{\star} \geq \min_{\mathbf{Z}}
\max_{\mathbf{s} \subseteq [\mathsf{N}], ~\text{subject to}~ \eqref{eq:vector constraint} } \frac{|\mathbf{s}|}{\mathsf{m}}-1. \label{eq:converse lemma}
\end{align}
\end{cor}

The essence of min-max problems lies in balancing extreme values across all chosen $\mathbf{s}$. 
 In order to simplify the computation, we provide a converse bound on $\eta^{\star}$, whose proof could be found in Appendix~\ref{sec:bound proof}.

\begin{cor}
\label{cor:converse cor}
For the $(\Ksf,\mathsf{N},\mathsf{N}_\mathrm{r}, \mathsf{m})$ secure gradient coding problem,
 it must hold that
\begin{align}
\eta^{\star} \geq \frac{\left\lceil  \frac{\mathsf{m}\mathsf{N}}{\mathsf{N}-\mathsf{N}_\mathrm{r}+\mathsf{m}} \right\rceil}{\mathsf{m}} -1. \label{eq:converse cor}
\end{align}
The proof could be found in Appendix~\ref{sec:cor2 proof}.
\end{cor}

Next, we propose a novel secure gradient coding scheme that employs a new data assignment strategy, improving the commonly used cyclic assignment.

\begin{figure*}
\centerline{\includegraphics[scale=0.21]{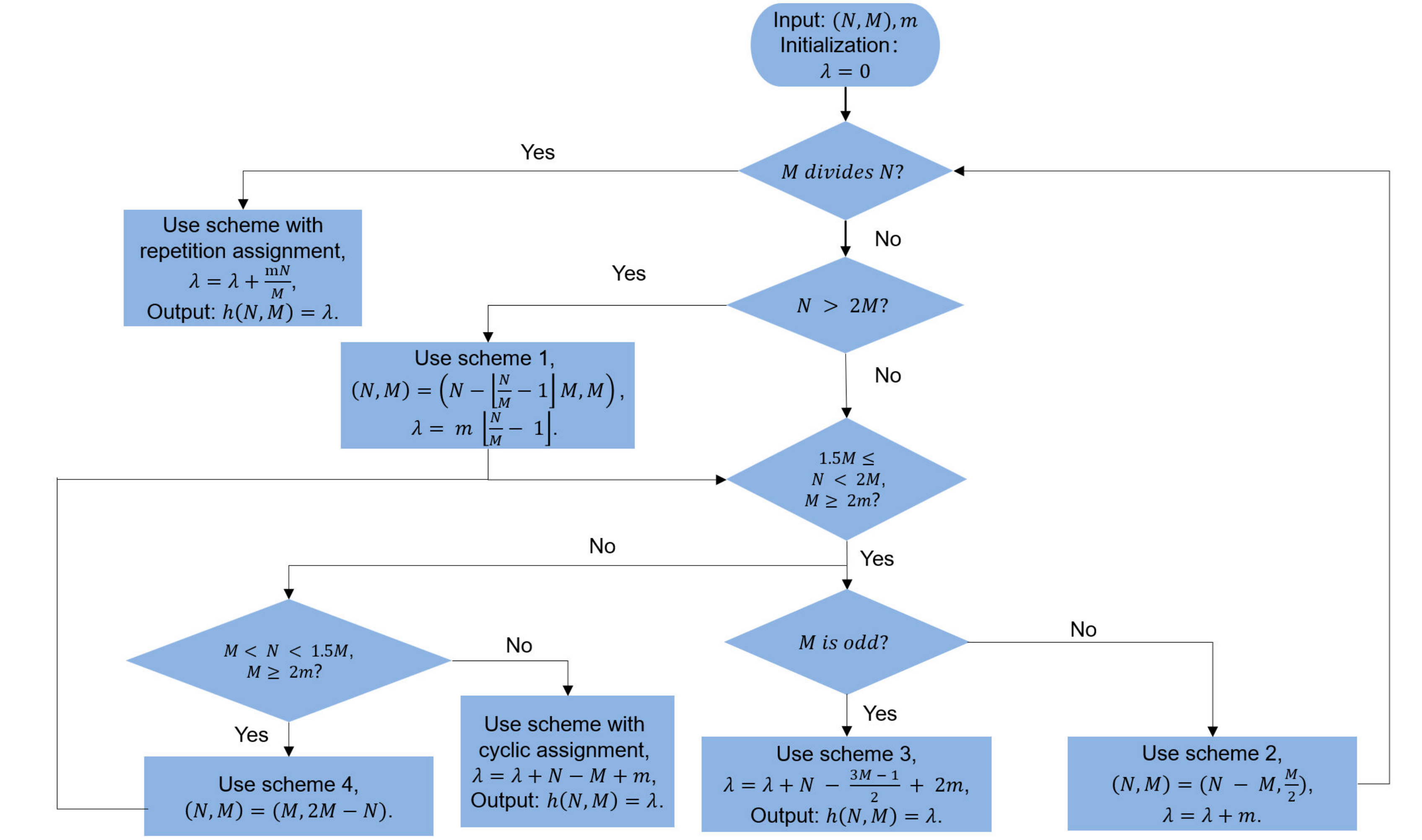}}
\caption{\small Flow diagram of the combined scheme in Theorem~\ref{thm:main achievable scheme}.}
\label{fig:scheme} 
\end{figure*}

\begin{thm}
\label{thm:main achievable scheme}
For the $(\Ksf,\mathsf{N},\mathsf{N}_\mathrm{r}, \mathsf{m})$ secure gradient coding problem, the source key size  $\eta=  \frac{ h(\mathsf{N},\mathsf{M})}{\mathsf{m}}-1$ is achievable, where the output of the function \( h(\cdot, \cdot) \) is given by the recursive algorithm shown in Fig.~\ref{fig:scheme} with the following properties:
 \begin{itemize}

 \item When  $\mathsf{N} > 2\mathsf{M}$,  by {\it Scheme~1} described in Section~\ref{sub:partial rep} we have  
\begin{align}
h(\mathsf{N},\mathsf{M})=h\big(\mathsf{N}-\left\lfloor \mathsf{N}/\mathsf{M}-1\right\rfloor \mathsf{M},\mathsf{M}\big)+ \mathsf{m}\left\lfloor \mathsf{N}/\mathsf{M} -1\right\rfloor  . \label{eq:from partial rep}
\end{align} 
\item When $1.5 \mathsf{M} \leq \mathsf{N} < 2 \mathsf{M}$ and $\mathsf{M}$ is even, with $\mathsf{M} \geq 2\mathsf{m}$,
by {\it Scheme~2} described in Section~\ref{sub:M is even}, we have  
\begin{align}
 h(\mathsf{N}, \mathsf{M}) = h\left(\mathsf{N} - \mathsf{M}, \frac{\mathsf{M}}{2}\right) + \mathsf{m}. \label{eq:M is even}
\end{align}

 \item When $1.5 \mathsf{M} \leq  \mathsf{N} < 2 \mathsf{M}$ and $\mathsf{M}$ is odd, with $\mathsf{M} \geq 2\mathsf{m}+1$, by  {\it Scheme~3} described in Section~\ref{sub:M is odd} we have  
 \begin{align}
 h(\mathsf{N},\mathsf{M})=   \mathsf{N}-\frac{3\mathsf{M}-1}{2}+2 \mathsf{m}. \label{eq:M is odd}
 \end{align}
 \item When $\mathsf{M}< \mathsf{N} <  1.5 \mathsf{M}$, with $\mathsf{M} \geq 2\mathsf{m}$, by  {\it Scheme~4} described in Section~\ref{sub:less than 1.5M}  we have  
\begin{align}
h(\mathsf{N},\mathsf{M})=h(\mathsf{M},2\mathsf{M}-\mathsf{N}). \label{eq:less than 1.5M}
\end{align} 
  \end{itemize}
    \hfill $\square$ 
 \end{thm}

\begin{rem}[The source key size in extreme cases]
\label{rem:extreme cases}
\em

For the $(\Ksf,\mathsf{N},\mathsf{N}_\mathrm{r}, \mathsf{m})$ secure gradient coding problem with  $\mathsf{M}= \mathsf{N} - \mathsf{N}_\mathrm{r} + \mathsf{m}$, and $\mathsf{M}$ divides $\mathsf{N}$, the minimum source key size is
\begin{align}
    \eta^{\star} = \frac{\mathsf{N}}{\mathsf{M}} - 1, \label{eq:gradient eta rho rep}
\end{align}
which matches the converse~\eqref{cor:converse cor} and is achievable using a fractional repetition assignment.

Under the cyclic assignment constraint, the minimum source key size required for optimal communication cost is
\begin{align}
    \eta^{\star}_{\text{cyc}} = \frac{\mathsf{N}_\mathrm{r}}{\mathsf{m}} - 1. \label{eq:gradient eta cyc}
\end{align}

\end{rem}

At the end of this section, we present numerical evaluations comparing the source key size in the converse in~\eqref{eq:converse cor}, in the cyclic assignment scheme, and the scheme from Theorem~\ref{thm:main achievable scheme}. 
The results demonstrate that the combined scheme requires a significantly smaller source key size than the cyclic assignment scheme and coincides with the converse at some points.

\begin{figure}[ht] 
    \centering
        \includegraphics[scale=0.4]{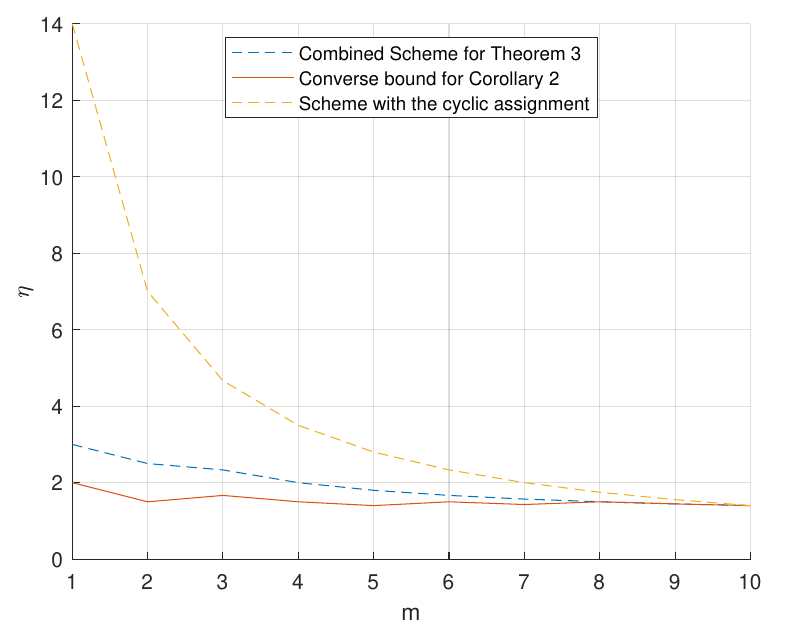}
    \caption{\small Numerical evaluations for the secure gradient coding with $\Ksf = \mathsf{N}=24, \mathsf{M}=10$.}
    \label{fig:numerical 1}
\end{figure}

\section{New Achievable Schemes For Theorem~\ref{thm:main achievable scheme}}
\label{sec:Achievable coding scheme}

For any secure distributed linearly separable scheme, \cite[Theorem 1]{wan2022secure} provides a transfer method from a non-secure scheme to a secure scheme while maintaining the communication cost.
The required source key size is $\frac{\lambda}{\mathsf{m}} - 1$, where $\lambda$ represents the number of linearly independent messages transmitted by all the servers, i.e., $H(X_1,\ldots, X_{\Ksf})/\Lsf$. Hence, to design a secure gradient coding scheme achieving the optimal communication cost with source key size as small as possible, we can design 
a non-secure gradient coding scheme achieving the optimal communication cost while minimizing the number of linearly independent messages.
Note that when $\mathsf{N} | \Ksf $, we can transfer the problem into $\Ksf = \mathsf{N}$ using~\cite[Remark 1]{wan2022secure}.  Thus, we focus on minimizing $\lambda$ in the non-secure scheme where $\Ksf= \mathsf{N}$. To elucidate the coding scheme, we present an illustrative example in Scheme~3, where the general description is given in Appendix~\ref{sec:general scheme}.

We mainly solve two challenges introduced by the transition from $\mathsf{m}=1$ to $\mathsf{m} > 1$: when dividing the servers into groups and assigning different datasets to different groups, datasets within the same group should be stored at least $\mathsf{m}$ times, we design a more complex structure; to reduce $\lambda$, we do not use the coding scheme in~\cite{Heterogeneous_Gradient_Coding}, but jointly design the coding scheme across different groups using transmitter Interference Alignment~\cite{2008Interference_Alignment, MIMO_interference_channel, huang2023ISITversion}.   
For details,  readers can refer to Appendix~\ref{sec:general scheme}.



\subsection{\texorpdfstring{Scheme~1 for~\eqref{eq:from partial rep}}{Scheme 1 for Eq. (X)}}
\label{sub:partial rep}
In this section, we consider the case where $\mathsf{N} > 2\mathsf{M}$. We partition servers and datasets into intervals of length \(\mathsf{M}\), i.e., \(\bigl[(i-1)\mathsf{M} + 1: i\mathsf{M}\bigr]\) for \(i \in \Bigl[\Bigl\lfloor \frac{\mathsf{N}}{\mathsf{M}} - 1 \Bigr\rfloor\Bigr]\), and assign them via the fractional repetition strategy, yielding \(\mathsf{m} \Bigl\lfloor \frac{\mathsf{N}}{\mathsf{M}} - 1 \Bigr\rfloor\) linearly independent combinations. We then assign the remaining \(\mathsf{N} - \Bigl\lfloor \frac{\mathsf{N}}{\mathsf{M}} - 1 \Bigr\rfloor \mathsf{M}\) datasets and servers, thereby reducing the original \((\mathsf{N}, \mathsf{M})\) problem to a smaller subproblem \(\bigl(\mathsf{N} - \bigl\lfloor \tfrac{\mathsf{N}}{\mathsf{M}} - 1 \bigr\rfloor \mathsf{M}, \mathsf{M}\bigr)\).

\subsection{\texorpdfstring{Scheme~2 for~\eqref{eq:M is even}}{Scheme 2 for Eq. (X)}}

\label{sub:M is even}

We consider the case \(1.5 \mathsf{M} \leq \mathsf{N} < 2 \mathsf{M}\) with even \(\mathsf{M} \geq 2\mathsf{m}\). The servers are divided into three groups: those in \([1, \tfrac{\mathsf{M}}{2}]\) (first group), those in \([\tfrac{\mathsf{M}}{2}+1, \mathsf{M}]\) (second group), and the remaining \(\mathsf{N} - \mathsf{M}\) servers (third group). The first two groups each provide \(\mathsf{m}\) linearly independent linear combinations, while the third group employs the \((\mathsf{N} - \mathsf{M}, \tfrac{\mathsf{M}}{2})\) scheme to recover \(h(\mathsf{N} - \mathsf{M}, \tfrac{\mathsf{M}}{2})\) combinations.

We first consider the data assignment for the servers in $[\mathsf{M}]$, whose structure is as follows.
\begin{align*}
\begin{array}{rl|c|c|c|c|c|c|c|}\cline{3-3}\cline{4-4}\cline{5-5}\cline{6-6}\cline{7-7}\cline{8-8}\cline{9-9}
&&\rule{0pt}{1.2em}\mbox{server}&\rule{0pt}{1.2em}\mbox{1} &\rule{0pt}{1.2em}\mbox{$\cdots$ } &  \rule{0pt}{1.2em}\mbox{ $\frac{\mathsf{M}}{2}$ } & \rule{0pt}{1.2em}\mbox{ $\frac{\mathsf{M}}{2}  +1$}&  \rule{0pt}{1.2em}\mbox{$\cdots$ } &  \rule{0pt}{1.2em}\mbox{$\mathsf{M}$ }\\ 
\cline{3-3}\cline{4-4}\cline{5-5}\cline{6-6}\cline{7-7}\cline{8-8}\cline{9-9}
&& &D_1 & \cdots & D_1 & D_1 & \cdots & D_1 \\
&& &\cdots & \cdots & \cdots & \cdots & \cdots & \cdots \\ 
&& \mbox{dataset}&D_{\sf y} & \cdots & D_{\sf y} & D_{\sf y} & \cdots & D_{\sf y} \\
&& &D_{\sf y+1} & \cdots & D_{\sf y+1} & D_{\mathsf{M}+1} & \cdots & D_{\mathsf{M}+1} \\ 
&& &\cdots & \cdots & \cdots & \cdots & \cdots & \cdots \\
&& &D_{\mathsf{M}} & \cdots & D_{\mathsf{M}}&  D_{\mathsf{N}} & \cdots & D_{\mathsf{N}}\\  
\cline{3-3}\cline{4-4}\cline{5-5}\cline{6-6}\cline{7-7}\cline{8-8}\cline{9-9}
\end{array}
\end{align*}

In this setup, we assign \(D_1\) to \(D_{\sf y}\) to all servers in \([\mathsf{M}]\), while each \(D_k\) for \(k \in [\sf y+1 : \mathsf{N}]\) is assigned to \(\frac{\mathsf{M}}{2}\) servers in \([\mathsf{M}]\). We then allocate the remaining \(\mathsf{N}-\mathsf{M}\) servers. Since \(\mathsf{N}-\sf y = 2(\mathsf{N}-\mathsf{M})\) datasets (from \([\sf y+1 : \mathsf{N}]\)) must be distributed among \(\mathsf{N}-\mathsf{M}\) servers, each dataset is assigned to \(\frac{\mathsf{M}}{2}\) servers, and each server receives \(\mathsf{M}\) datasets. 

To implement this, we divide the datasets in \([\sf y+1 : \mathsf{N}]\) into \(\frac{\mathsf{N}-\sf y}{2} = \mathsf{N}-\mathsf{M}\) pairs \(\Pc_i = \{\sf y + i, \mathsf{M} + i\}\) for \(i \in [\mathsf{N}-\mathsf{M}]\). We then use the assignment phase of the \(\bigl(\mathsf{N}-\mathsf{M}, \tfrac{\mathsf{M}}{2}\bigr)\) scheme, assigning each pair to \(\tfrac{\mathsf{M}}{2}\) servers, with each server receiving \(\tfrac{\mathsf{M}}{2}\) pairs.

\subsection{\texorpdfstring{Scheme~3 for~\eqref{eq:M is odd}}{Scheme 3 for Eq. (Y)}}
\label{sub:M is odd}
We provide an example to illustrate the main idea.
\begin{example} \rm
\label{ex:scheme 3 example}
We consider the $(\mathsf{N}, \mathsf{M}) = (12, 7)$ non-secure problem, where $\mathsf{m} = 2$.

{\it Data assignment phase.}
We assign the datasets as follows.

\begin{align*}
{\small 
\begin{array}{rl|c|c|c|}\cline{3-3}\cline{4-4}\cline{5-5}
&&\rule{0pt}{1.2em}\mbox{server}&\rule{0pt}{1.2em}\mbox{\textnormal{1}}  &\rule{0pt}{1.2em}\mbox{\textnormal{2}} \\ \cline{3-3}\cline{4-4}\cline{5-5}
&& &D_1 & D_1 \\
&& &D_2 & D_2  \\
&& &D_3 & D_3 \\
&& \mbox{dataset}&D_4 & D_4  \\
&& &D_5 & D_5 \\
&& &D_6 & D_6 \\
&& &D_7 & D_7\\
\cline{3-3}\cline{4-4}\cline{5-5}
\end{array}
\hspace{-0.5em} 
\begin{array}{rl|c|c|c|c|c|}\cline{3-3}\cline{4-4}\cline{5-5}\cline{6-6}\cline{7-7}
&&\rule{0pt}{1.2em}\mbox{\textnormal{3}} & \rule{0pt}{1.2em}\mbox{\textnormal{4}} & \rule{0pt}{1.2em}\mbox{\textnormal{5}} & \rule{0pt}{1.2em}\mbox{\textnormal{6}} & \rule{0pt}{1.2em}\mbox{\textnormal{7}} \\ \cline{3-3}\cline{4-4}\cline{5-5}\cline{6-6}\cline{7-7}
&&  D_1 & D_1 & D_1 & D_1 & D_1\\
&&   D_2 & D_2 & D_2 & D_2 & D_2 \\
&&  D_3 & D_3 & D_3 & D_3 & D_3\\
&&  D_8 & D_9 & D_{10} & D_{11} & D_{12} \\
&&  D_9 & D_{10} & D_{11} & D_{12} & D_8\\
&&  D_{10} & D_{11} & D_{12} & D_8 & D_9\\
&&  D_{11} & D_{12} & D_8 & D_9 & D_{10}\\
\cline{3-3}\cline{4-4}\cline{5-5}\cline{6-6}\cline{7-7}
\end{array}
}
\end{align*}

\begin{align*}
\begin{array}{rl|c|c|c|c|c|}\cline{3-3}\cline{4-4}\cline{5-5}\cline{6-6}\cline{7-7}
&&\rule{0pt}{1.2em}\mbox{\textnormal{8}} & \rule{0pt}{1.2em}\mbox{\textnormal{9}}  & \rule{0pt}{1.2em}\mbox{\textnormal{10}} & \rule{0pt}{1.2em}\mbox{\textnormal{11}} & \rule{0pt}{1.2em}\mbox{\textnormal{12}} \\ \cline{3-3}\cline{4-4}\cline{5-5}\cline{6-6}\cline{7-7}
&&  D_4& D_4 & D_4 & D_4 & D_4\\
&&  D_5& D_5 & D_5 & D_5 & D_5\\
&&  D_6& D_6 & D_6 & D_6 & D_6\\
&&  D_7& D_7 & D_7 & D_7 & D_7\\
&&  D_8& D_9 & D_{10} & D_{11} & D_{12}\\
&&  D_9& D_{10} & D_{11} & D_{12} & D_8\\
&&  D_{10}& D_{11} & D_{12} & D_8 & D_9 \\
\cline{3-3}\cline{4-4}\cline{5-5}\cline{6-6}\cline{7-7}
\end{array}
\end{align*}

{\it Computing phase.}
To minimize communication cost, each message $W_k$, for $k \in [\Ksf]$, is divided into $\mathsf{m}$ equal-length sub-messages, denoted as $W_k = \{W_{k,j} : j \in [\mathsf{m}]\}$, with each sub-message containing $\frac{\Lsf}{\mathsf{m}} = \frac{\Lsf}{2}$ symbols in $\mathbb{F}_{\qsf}$. Each server sends a linear combination of messages, allowing the user to recover
\(\mathbf{F} [W_{1,1}; \ldots; W_{12,2}]\) from any \(\mathsf{N}_\mathrm{r} = 7\) servers, where
\(\mathbf{F}\) is shown at the top of the page in \eqref{eq:F of example3}. The design of \(\mathbf{f}_5\) and \(\mathbf{f}_6\) will be detailed later.
We define that $[F_1;F_2;F_3;F_4;F_5;F_6]=  {\bf F} \  [W_{1,1};\ldots;W_{12,2}] $.
\begin{figure*}
\begin{equation}
\scalebox{0.8}{%
\(
{\bf F} = \begin{bmatrix}  
 {\mathbf{f}}_1 \\
 {\mathbf{f}}_2 \\
 {\mathbf{f}}_3 \\
 {\mathbf{f}}_4\\
 {\mathbf{f}}_5 \\
 {\mathbf{f}}_6
\end{bmatrix}
=
\left[
\begin{array}{cccccccccccccccccccccccc}
  1 & 1 & 1 & 1 & 1 & 1 & 1 & 1 & 1 & 1 & 1 & 1 & 0 & 0 & 0 & 0 & 0 & 0 & 0 & 0 & 0 & 0 & 0 & 0 \\
  0 & 0 & 0 & 0 & 0 & 0 & 0 & 0 & 0 & 0 & 0 & 0 & 1 & 1 & 1 & 1 & 1 & 1 & 1 & 1 & 1 & 1 & 1 & 1 \\
  0 & 0 & 0 & 2 & 2 & 2 & 2 & 1 & 1 & 1 & 1 & 1 & 0 & 0 & 0 & 0 & 0 & 0 & 0 & 0 & 0 & 0 & 0 & 0 \\
  0 & 0 & 0 & 0 & 0 & 0 & 0 & 0 & 0 & 0 & 0 & 0 & 0 & 0 & 0 & 2 & 2 & 2 & 2 & 1 & 1 & 1 & 1 & 1 \\
  0 & 0 & 0 & 0 & 0 & 0 & 0 & 2 & 4 & 2 & 4 & 2 & 0 & 0 & 0 & 0 & 0 & 0 & 0 & 2 & 1 & 3 & 4 & 2 \\
  0 & 0 & 0 & 0 & 0 & 0 & 0 & 2 & 2 & 4 & 4 & -2 & 0 & 0 & 0 & 0 & 0 & 0 & 0 & 3 & 3 & 1 & 1 & -5 \\
\end{array}
\right]
\)
}
\label{eq:F of example3}
\end{equation}
\end{figure*}

\paragraph*{Key step}
\textit{We divide the servers into three groups based on the data assignment: servers 1 and 2 form the first group, those in $[3\!:\!7]$ form the second, and those in $[8\!:\!12]$ form the third. We require that the $\mathsf{m}=2$ linear combinations \((F_1 - F_3)\) and \((F_2 - F_4)\) be recovered from the first group. From the second group, we recover \((2F_1 - F_3)\), \((2F_2 - F_4)\), \(F_5\), and \(F_6\), which together form \(\mathsf{N} - \frac{3\mathsf{M} - 1}{2} + \mathsf{m} = 4\) linear combinations, and from the third group, we recover \(F_3\), \(F_4\), \(F_5\), and \(F_6\), another \(4\) combinations. Hence, we can recover 
$\mathsf{N} - \frac{3\mathsf{M} - 1}{2} + 2\mathsf{m} = 6$ linear combinations in total, matching the result in~\eqref{eq:M is odd}. The following outlines the specific procedure.}


We focus on servers 1 and 2, which share the same datasets and exclude those from $[8{:}11]$. 
server~1 computes 
\[
X_1 = (F_1 - F_3) + (F_2 - F_4),
\]
and server~2 computes
\[
X_2 = (F_1 - F_3) - (F_2 - F_4).
\]
For servers handling $[8{:}12]$, we design transmissions so that $F_3, F_4, F_5,$ and $F_6$ 
can be recovered from any four responses. Specifically, for $n \in [8{:}12]$, let 
server~$n$ compute
\[
\mathbf{s}_n \bigl[ \mathbf{f}_3 ; \mathbf{f}_4 ; \mathbf{f}_5 ; \mathbf{f}_6 \bigr] 
\bigl[ W_{1,1}; \ldots; W_{12,2} \bigr].
\]

We design ${\bf s}_n$ according to the following steps. We focus on the datasets in $[8:12]$ under cyclic assignment, extracting the corresponding columns from the matrix to form a submatrix ${\bf F}^{\prime}_1$. 

Notice that server $n$ cannot compute two datasets in $[8:12]$. We extract the corresponding columns from the matrix to form a submatrix $\overline{{\bf F}^{\prime}_1}$.

Our required ${\bf s}_n$ is the left null vector of $\overline{{\bf F}^{\prime}_1}$. 
There should be a linear combination of the columns in $\overline{{\bf F}^{\prime}_1}$ leading to one rank deficiency, with the form
\begin{align}
 {\bf F}'_1 \mathbf{e}_n^{\text{\rm T}} = {\bf 0}_{4 \times 1}.
 \label{eq: linear reduction equation}
\end{align}
For the first row of $\overline{{\bf F}^{\prime}_1}$, since $\left[ {\begin{array}{cc} 1 & 1 \\ \end{array}} \right] [1, -1]^{\text{\rm T}} = {\bf 0}_{1 \times 1}$, the corresponding elements in $\mathbf{e}_n$ should be a multiple of $(1, -1)$, and we choose a random multiple. Similarly, the same applies to the second row of $\overline{{\bf F}^{\prime}_1}$. Repeating the above steps, we obtain
the matrix ${\bf E}$, where
\begin{align}
{\bf E} = 
\scalebox{0.8}{$
\begin{bmatrix}
\mathbf{e}_1\\
\vdots \\
\mathbf{e}_5
\end{bmatrix} =  
\begin{bmatrix}
0 & 0 & 0 & 1 & -1 & 0 & 0 & 0 & -1 & 1 \\
2 & 0 & 0 & 0 & -2 & -1 & 0 & 0 & 0 & 1 \\
1 & -1 & 0 & 0 & 0 & 2 & -2 & 0 & 0 & 0 \\
0 & 1 & -1 & 0 & 0 & 0 & 1 & -1 & 0 & 0 \\
0 & 0 & -1 & 1 & 0 & 0 & 0 & 2 & -2 & 0 \\
\end{bmatrix}
$}
\label{eq:E for example}
\end{align}


By \eqref{eq: linear reduction equation} and \eqref{eq:E for example}, we have
\begin{align}
{\bf F}'_1 {\bf E}^{\text{\rm T}} = \mathbf{0}_{4\times 5}, 
\label{eq:equations for example}
\end{align}
where ${\bf E}^{\text{\rm T}}$ is \(10 \times 5\) and full rank. Thus, its left null space contains \(5\) linearly independent vectors, spanning the rows of ${\bf F}'_1$. The first two rows lie in the null space of ${\bf E}^{\text{\rm T}}$, and any two vectors from the remaining three null space vectors complete ${\bf F}'_1$, given as:
\[
{\bf F}'_1 = 
\scalebox{0.8}{$
\begin{bmatrix}
1 & 1 & 1 & 1 & 1 & 0 & 0 & 0 & 0 & 0 \\
0 & 0 & 0 & 0 & 0 & 1 & 1 & 1 & 1 & 1 \\
2 & 4 & 2 & 4 & 2 & 2 & 1 & 3 & 4 & 2 \\
2 & 2 & 4 & 4 & -2 & 3 & 3 & 1 & 1 & -5
\end{bmatrix}
$}.
\]
For \({\bf s}_n\), satisfying \({\bf s}_n \overline{{\bf F}'_1} = \mathbf{0}_{1 \times 4}\), the resulting matrix is:
\[
{\bf S} = 
\begin{bmatrix}
{\bf s}_{1} \\
\vdots \\
{\bf s}_{5}
\end{bmatrix} =  
\scalebox{0.8}{$
\begin{bmatrix}
 2 & 11/4 & -3/4 & 1/4 \\
 4 & 4 & -2 & 0 \\
 2 & 3 & 0 & -1 \\
 8 & 16/3 & -4/3 & -4/3 \\
 6 & 3/2 & 0 & -3/2 \\
\end{bmatrix}
$}.
\]
For \(n \in [8, 12]\), server~\(n\) computes 
\[
X_n = {\bf s}_n \bigl[F_3; F_4; F_5; F_6\bigr].
\]

Next, servers in \([3{:}7]\), excluding datasets from \([4{:}7]\), ensure that responses from any \(4\) servers recover \((2F_1 - F_3)\), \((2F_2 - F_4)\), \(F_5\), and \(F_6\). For \(n \in [3, 7]\), server~\(n\) computes
\[
X_n = {\bf s}_n \bigl[2{\mathbf{f}}_1 - {\mathbf{f}}_3; 2{\mathbf{f}}_2 - {\mathbf{f}}_4; {\mathbf{f}}_5; {\mathbf{f}}_6\bigr]\bigl[W_{1,1}; \ldots; W_{12,2}\bigr].
\]

For datasets in \([8{:}12]\), cyclic assignment forms \({\bf F}'_2\), and it is evident that \({\bf F}'_2 = {\bf F}'_1\). Since datasets missing from server~\(n \in [3{:}7]\) are included in those missing from \(n+5\), we have \({\bf S}' = {\bf S}\). Thus, server~\(n\) computes
\[
X_n = {\bf s}_n \bigl[2F_1 - F_3; 2F_2 - F_4; F_5; F_6\bigr].
\]

In conclusion, under the condition $\mathsf{m} = 2$, the number of totally transmitted linearly independent combinations is $h(12,7) = 12 - 10 + 4 = 6$, which coincides with~\eqref{eq:M is odd}. Moreover, our calculations show that if
\[
\mathsf{m}(\mathsf{N} - \mathsf{M} - 1) \;\leq\; \bigl[(\mathsf{N} - \mathsf{M})(\mathsf{m} - 1) + 1\bigr]\Bigl(\mathsf{N} - \tfrac{3}{2}\mathsf{M} - \tfrac{1}{2}\Bigr),
\]
transmitter Interference Alignment~\cite{2008Interference_Alignment, MIMO_interference_channel, huang2023ISITversion} is required.

\end{example}

\subsection{\texorpdfstring{Scheme~4 for~\eqref{eq:less than 1.5M}}{Scheme 4 for Eq. (Z)}}
\label{sub:less than 1.5M}
In this section, we consider the case \( \mathsf{M} < \mathsf{N} < 1.5 \mathsf{M} \) with \( \mathsf{M} \ge 2 \mathsf{m} \) and transform the \((\mathsf{N},\mathsf{M})\) problem into \((\mathsf{M}, 2\mathsf{M} - \mathsf{N})\) by dividing the servers into two groups: those in \([\mathsf{M}]\) and those in \([\mathsf{M}+1 : \mathsf{N}]\).
We assign \(D_1, \ldots, D_{\mathsf{N} - \mathsf{M}}\) to each server in \([\mathsf{M}]\), and \(D_{\mathsf{N} - \mathsf{M} + 1}, \ldots, D_{\mathsf{N}}\) to each server in \([\mathsf{M}+1 : \mathsf{N}]\). Thus, every server in \([\mathsf{M}+1 : \mathsf{N}]\) holds \(\mathsf{M}\) datasets, while each server in \([\mathsf{M}]\) has \(\mathsf{N} - \mathsf{M}\). Since every dataset in \([\mathsf{N} - \mathsf{M} + 1 : \mathsf{N}]\) is assigned to \(\mathsf{N} - \mathsf{M}\) servers, we further assign each \(D_k\) (for \(k \in [\mathsf{N} - \mathsf{M} + 1 : \mathsf{N}]\)) to \(2\mathsf{M} - \mathsf{N}\) additional servers in \([\mathsf{M}]\), giving each such server \((2\mathsf{M} - \mathsf{N})\) datasets. This allocation follows the assignment phase of the \((\mathsf{M}, 2\mathsf{M} - \mathsf{N})\) non-secure problem.

\clearpage
\bibliographystyle{IEEEtran}
\bibliography{re} 
\clearpage

  \appendices

\section{Proof of Theorem~\ref{thm:bound of η}}
\label{sec:bound proof}
By the security constraint in \eqref{eq:security}, the user can only obtain $W_1+\cdots+W_{\Ksf}$ without accessing any additional information about the messages, even after receiving answers from all servers. Let $X_{\Sc}=\{X_n: n \in \Sc\}$. From~\eqref{eq:security}, we derive:
\begin{subequations}
\begin{align}
0 &= I(W_1, \ldots, W_{\Ksf}; X_{[\mathsf{N}]} | W_1+W_2+\cdots+W_{\Ksf}) \\
  &= H(X_{[\mathsf{N}]} | W_1+W_2+\cdots+W_{\Ksf}) - H(X_{[\mathsf{N}]} | W_1, \ldots, W_{\Ksf}) \\
  &\geq H(X_{[\mathsf{N}]} | W_1+W_2+\cdots+W_{\Ksf}) \nonumber\\
  &- H(Q, W_1, \ldots, W_{\Ksf} | W_1, \ldots, W_{\Ksf}) \label{eq:function of QW} \\
  &= H(X_{[\mathsf{N}]} | W_1+W_2+\cdots+W_{\Ksf}) - H(Q) \label{eq:independent noise} \\
  &= H(X_{[\mathsf{N}]}) - I(X_{[\mathsf{N}]}; W_1+W_2+\cdots+W_{\Ksf}) - H(Q) \\
  &\geq H(X_{[\mathsf{N}]}) - H(W_1+W_2+\cdots+W_{\Ksf}) - H(Q), \label{eq:shannon}
\end{align}
\end{subequations}
where~\eqref{eq:function of QW} follows from the fact that $X_{[\mathsf{N}]}$ is a function of $Q$ and $W_1, \ldots, W_{\Ksf}$, and~\eqref{eq:independent noise} relies on $Q$ being independent of $W_1, \ldots, W_{\Ksf}$.

From~\eqref{eq:shannon}, we further obtain:
\begin{subequations}
\begin{align}
\eta \Lsf &\geq H(Q) \geq H(X_{[\mathsf{N}]}) - H(W_1+\cdots+W_{\Ksf}) \\
          &\geq H(X_{[\mathsf{N}]}) - \Lsf \label{eq:W1 inde} \\
          &\geq H(X_{s_1}, \ldots, X_{s_{|\mathbf{s}|}}) - \Lsf \\
          &\geq  \frac{|\mathbf{s}|\Lsf}{\mathsf{m}} - \Lsf, \label{eq:lemma converse all term}
\end{align}
\end{subequations}
where~\eqref{eq:W1 inde} follows from the independence of $\Ksf$ messages, each uniformly i.i.d. over $[\mathbb{F}_{\qsf}]^{\Lsf}$. 

Next, I will prove~\eqref{eq:lemma converse all term} under the symmetric transmission and linear coding. Note that server $s_i$ always contains at least one message $W_{s_i}$ which can not be computed by $m$ servers from $\{s_1, \ldots, s_{i-1}\}$.  Consider the case that apart from $\mathsf{N} - \mathsf{N}_\mathrm{r}$ stragglers, there are exactly $m$ servers who can compute $W_{s_i}$. Under the symmetric transmission and linear coding, each server should contain $\frac{1}{m}$ independent part of $W_{s_i}$. Thus, we have the message of server $s_i$ is independent to messages from $\{s_1, \ldots, s_{i-1}\},$ because of the independent part of $W_{s_i}$. Since the length of each message should be larger than $\frac{\Lsf}{\mathsf{m}}$, we can obtain~\eqref{eq:lemma converse all term} under the symmetric transmission and linear coding. 

\section{Proof of Corollary~\ref{cor:converse cor}}
\label{sec:cor2 proof} 
By definition, there are $\Ksf$ datasets in the library and we assign $\frac{\Ksf}{\mathsf{N}}(\mathsf{N}-\mathsf{N}_\mathrm{r}+\mathsf{m})$ datasets to
each server.   
Hence, for any possible assignment, we can find 
$$
\left\lceil \frac{\mathsf{m}\Ksf}{\mathsf{M}} \right\rceil= \left\lceil \frac{\mathsf{m}\mathsf{N}}{\mathsf{N}-\mathsf{N}_\mathrm{r}+\mathsf{m}} \right\rceil
$$ servers, where each server has some dataset which appears at most $\mathsf{m}-1$ times in other $\left\lceil \frac{\mathsf{m}\mathsf{N}}{\mathsf{N}-\mathsf{N}_\mathrm{r}+\mathsf{m}} \right\rceil-1$ servers.  By Theorem~\ref{thm:bound of η}, we have 
 $\eta^{\star} \geq \frac{\left\lceil  \frac{\mathsf{m}\mathsf{N}}{\mathsf{N}-\mathsf{N}_\mathrm{r}+\mathsf{m}} \right\rceil}{\mathsf{m}} -1 $.

\section{The general scheme of Theorem~\ref{thm:main achievable scheme}}
\label{sec:general scheme} 
\subsection{general scheme 1}
\label{general scheme 1}
{\it Data assignment phase.}
We partition the system into $\left\lfloor  \mathsf{N}/\mathsf{M} \right\rfloor $ blocks. For each $i \in [\left\lfloor \mathsf{N}/\mathsf{M} \right\rfloor - 1]$, the $i^{\text{th}}$ block includes the datasets $\{ D_{k}: k \in \left[(i-1)\mathsf{M}+1: i\mathsf{M} \right]\}$ and the servers in the range $\left[(i-1)\mathsf{M}+1: i\mathsf{M} \right]$.
The last block includes the datasets and servers in the $\left[\left\lfloor \mathsf{N}/\mathsf{M} - 1 \right\rfloor \mathsf{M} + 1: \mathsf{N} \right]$.

We use the fractional repetition assignment to distribute the data for the first $\left[\left\lfloor \mathsf{N}/\mathsf{M} \right\rfloor - 1\right]$ blocks. As for the last block, it contains $\mathsf{N} - \left\lfloor \mathsf{N}/\mathsf{M} - 1 \right\rfloor \mathsf{M}$ servers and $\mathsf{N} - \left\lfloor \mathsf{N}/\mathsf{M} - 1 \right\rfloor \mathsf{M}$ datasets, with each server being able to compute $M$ datasets. Therefore, we can directly apply the existing data assignment scheme designed for solving the $(\mathsf{N} - \left\lfloor \mathsf{N}/\mathsf{M} - 1 \right\rfloor \mathsf{M}, \mathsf{M})$ non-secure problem to this last block.

{\it Computing phase.}
For each $i \in \left[\left\lfloor \mathsf{N}/\mathsf{M} \right\rfloor - 1\right]$, each server in the $i^{\text{th}}$ block computes $m$ linear combinations $\sum_{k\in \left[(i-1)\mathsf{M} + 1 : i\mathsf{M}\right]} W_{k,j}$,  and each server sends a random linear combination of these $m$ computed combinations.

Next, we focus on the last block. The computation phase in this block still follows the $(\mathsf{N} - \left\lfloor \mathsf{N}/\mathsf{M} - 1 \right\rfloor \mathsf{M}, \mathsf{M})$ non-secure scheme, which results in a total of $h\left(\mathsf{N} - \left\lfloor \mathsf{N}/\mathsf{M} - 1 \right\rfloor \mathsf{M}, \mathsf{M}\right)$ linearly independent combinations being sent.

{\it Decoding phase.}
The user receives information from $\mathsf{N}_\mathrm{r} = \mathsf{N} - \mathsf{M} + \mathsf{m}$ servers, meaning that responses from $\mathsf{M} - \mathsf{m}$ servers will not be received. For the first $\left\lfloor \mathsf{N}/\mathsf{M} - 1 \right\rfloor$ blocks, each block has $\mathsf{M}$ servers, so the user receives at least $\mathsf{m}$ servers' responses from each block. Thus, from each block, the user can recover $\sum_{k \in \left[(i-1)\mathsf{M} + 1 : i\mathsf{M}\right]} W_{k,j}$, where $j$ ranges over $[\mathsf{m}]$. 
For the last block, at least $\mathsf{N} - \left\lfloor \mathsf{N}/\mathsf{M} - 1 \right\rfloor \mathsf{M} - \mathsf{M} + \mathsf{m}$ servers will respond. By carefully designing the coding scheme, we can recover $\sum_{k \in \left[\left\lfloor \mathsf{N}/\mathsf{M} - 1 \right\rfloor \mathsf{M} + 1 : \mathsf{N}\right]} W_k$ from the information sent by any $\mathsf{N} - \left\lfloor \mathsf{N}/\mathsf{M} - 1 \right\rfloor \mathsf{M} - \mathsf{M} + \mathsf{m}$ servers in the last block.
together with the transmissions of the first $\left\lfloor  \mathsf{N}/\mathsf{M} -1 \right\rfloor$ blocks, the user can recover $W_1 + \cdots +W_{\mathsf{N}}$.

In conclusion, we have demonstrated that $h(\mathsf{N}, \mathsf{M}) = \mathsf{m}\left\lfloor \mathsf{N}/\mathsf{M} - 1 \right\rfloor + h\left(\mathsf{N} - \left\lfloor \mathsf{N}/\mathsf{M} - 1 \right\rfloor \mathsf{M}, \mathsf{M} \right)$, which aligns with equation~\eqref{eq:from partial rep}. 
Furthermore, it is evident that $\mathsf{M} < \mathsf{N} - \left\lfloor \mathsf{N}/\mathsf{M} - 1 \right\rfloor \mathsf{M} < 2\mathsf{M}$ when $\mathsf{N} > 2\mathsf{M}$ and $\mathsf{M}$ does not evenly divide $\mathsf{N}$.

\subsection{general scheme 2}
\label{general scheme 2}
We now focus on the case where $1.5 \mathsf{M} \leq \mathsf{N} < 2 \mathsf{M}$, and $\mathsf{M}$ is even, with $\mathsf{M} \geq 2\mathsf{m}$. We consider the general Scheme 2 for the $(\mathsf{N}, \mathsf{M})$ non-secure problem to prove~\eqref{eq:M is even}. We assume that a specific feasible scheme already exists for the $\left(\mathsf{N} - \mathsf{M}, \frac{\mathsf{M}}{2}\right)$ non-secure problem and that this scheme can transmit a total of $h\left(\mathsf{N} - \mathsf{M}, \frac{\mathsf{M}}{2}\right)$ linearly independent combinations of messages.

{\it Data assignment phase.}
We first consider the dataset assignment for the servers in $[\mathsf{M}]$, which is structured as follows:
\begin{align*}
\begin{array}{rl|c|c|c|c|c|c|c|}\cline{3-3}\cline{4-4}\cline{5-5}\cline{6-6}\cline{7-7}\cline{8-8}\cline{9-9}
&&\rule{0pt}{1.2em}\mbox{server} & \rule{0pt}{1.2em}\mbox{1} &\rule{0pt}{1.2em}\mbox{$\cdots$ } &  \rule{0pt}{1.2em}\mbox{ $\frac{\mathsf{M}}{2}$ } & \rule{0pt}{1.2em}\mbox{ $\frac{\mathsf{M}}{2}  +1$}&  \rule{0pt}{1.2em}\mbox{$\cdots$ } &  \rule{0pt}{1.2em}\mbox{$\mathsf{M}$ }\\ 
\cline{3-3}\cline{4-4}\cline{5-5}\cline{6-6}\cline{7-7}\cline{8-8}\cline{9-9}
&& &D_1 & \cdots & D_1 & D_1 & \cdots & D_1 \\
&& &\cdots & \cdots & \cdots & \cdots & \cdots & \cdots \\ 
&& \mbox{dataset}&D_{\sf y} & \cdots & D_{\sf y} & D_{\sf y} & \cdots & D_{\sf y} \\
&& &D_{\sf y+1} & \cdots & D_{\sf y+1} & D_{\mathsf{M}+1} & \cdots & D_{\mathsf{M}+1} \\ 
&& &\cdots & \cdots & \cdots & \cdots & \cdots & \cdots \\
&& &D_{\mathsf{M}} & \cdots & D_{\mathsf{M}}&  D_{\mathsf{N}} & \cdots & D_{\mathsf{N}}\\  
\cline{3-3}\cline{4-4}\cline{5-5}\cline{6-6}\cline{7-7}\cline{8-8}\cline{9-9}
\end{array}
\end{align*}

In this setup, we assign datasets $D_1$ through $D_{\sf y}$ to all servers in $[\mathsf{M}]$, while each dataset $D_k$ for $k \in [\sf y+1 : \mathsf{N}]$ is distributed among $\frac{\mathsf{M}}{2}$ servers in $[\mathsf{M}]$.

Next, we consider the assignment for the last $\mathsf{N}-\mathsf{M}$ servers.
We need to allocate $\mathsf{N}-\sf y = 2(\mathsf{N}-\mathsf{M})$ datasets (from $[\sf y+1 : \mathsf{N}]$) to a total of $\mathsf{N}-\mathsf{M}$ servers. Each dataset is assigned to $\frac{\mathsf{M}}{2}$ servers, and each server receives $\mathsf{M}$ datasets. 

Datasets in $[\sf y+1 : \mathsf{N}]$ are divided into $\frac{\mathsf{N}-\sf y}{2} = \mathsf{N}-\mathsf{M}$ pairs, where the $i^{\text{th}}$ pair is defined as $\Pc_i = \{\sf y+i, \mathsf{M}+i\}$ for $i \in [\mathsf{N}-\mathsf{M}]$. Thus, we apply the assignment phase of the scheme for the $\left(\mathsf{N}-\mathsf{M}, \frac{\mathsf{M}}{2}\right)$ non-secure problem, assigning each pair to $\frac{\mathsf{M}}{2}$ servers, with each server being allocated $\frac{\mathsf{M}}{2}$ pairs.

{\it Computing phase.}
We divide each message $W_k$ for $k \in [\Ksf]$ into $\mathsf{m}$ equal-length, non-overlapping sub-messages, denoted as $W_k = \{W_{k,j} : j \in [\mathsf{m}]\}$.
We first focus on the servers in $[\mathsf{M}]$. By constructing the messages sent by these servers, we can recover the following $2\mathsf{m}$ linear combinations:
\begin{align*}
    F_1 &= W_{1,1} + W_{2,1} + \cdots + W_{\mathsf{N},1}, \\
    &\vdots \\
    F_{\mathsf{m}} &= W_{1,\mathsf{m}} + W_{2,\mathsf{m}} + \cdots + W_{\mathsf{N},\mathsf{m}}, \\
    \\
    F_{\mathsf{m}+1} &= 2(W_{\sf y+1,1} + \cdots + W_{\mathsf{M},1}) + W_{\mathsf{M}+1,1} + \cdots + W_{\mathsf{N},1}, \\
    &\vdots \\
    F_{2\mathsf{m}} &= 2(W_{\sf y+1,\mathsf{m}} + \cdots + W_{\mathsf{M},\mathsf{m}}) + W_{\mathsf{M}+1,\mathsf{m}} + \cdots + W_{\mathsf{N},\mathsf{m}}.
\end{align*}

We group the $\mathsf{m}$ linear combinations $(F_1 - F_{\mathsf{m}+1}), \ldots, (F_{\mathsf{m}} - F_{2\mathsf{m}})$ into one group, and the $\mathsf{m}$ linear combinations $(2F_1 - F_{\mathsf{m}+1}), \ldots, (2F_{\mathsf{m}} - F_{2\mathsf{m}})$ into another group. Since the servers in $[\mathsf{M}/2]$ are not assigned any data from $[\mathsf{M}+1: \mathsf{N}]$, we let each server in this range compute a random linear combination of the $\mathsf{m}$ linear combinations in the first group. The servers in $[\mathsf{M}/2+1: \mathsf{M}]$, who are not assigned any data from $[\sf y+1: \mathsf{M}]$, are assigned to compute a random linear combination of the $\mathsf{m}$ linear combinations in the second group.

We then focus on the servers in $[\mathsf{M}+1: \mathsf{N}]$. For each dataset pair $\Pc_i = \{\sf y+i, \mathsf{M}+i\}$ where $i \in [\mathsf{N} - \mathsf{M}]$, we define $P_i = 2W_{\sf y+i} + W_{\mathsf{M}+i}$. Therefore, we can express $F_{\mathsf{m}+1}$ to $F_{2\mathsf{m}}$ as $P_{1,1} + \cdots + P_{\mathsf{N}-\mathsf{M},1}, \ldots, P_{1,\mathsf{m}} + \cdots + P_{\mathsf{N}-\mathsf{M},\mathsf{m}}$.
Thus, we can apply the computing phase of the proposed scheme to the $\left(\mathsf{N} - \mathsf{M}, \frac{\mathsf{M}}{2}\right)$ non-secure problem.
\\

{\it Decoding phase.}

The user receives $\mathsf{N}_\mathrm{r} = \mathsf{N} - \mathsf{M} + \mathsf{m}$ server responses, which can be categorized into two cases.

First, we consider the case where at least $\mathsf{N} - 1.5\mathsf{M} + \mathsf{m}$ responses come from servers in $[\mathsf{M}+1 : \mathsf{N}]$. In this scenario, $F_{\mathsf{m}+1}$ to $F_{2\mathsf{m}}$ can be recovered. Additionally, at least $\mathsf{m}$ responses from servers in $[\mathsf{M}]$ are received, providing at least $\mathsf{m}$ linearly independent combinations. Together with the combinations from $F_{\mathsf{m}+1}$ to $F_{2\mathsf{m}}$, this gives a total of $2\mathsf{m}$ linearly independent combinations, enabling recovery of $F_1$ to $F_{2\mathsf{m}}$.
We then consider the second case, where the servers in $[\mathsf{M}+1: \mathsf{N}]$ respond with $\mathsf{N} - 1.5\mathsf{M} + \mathsf{m} - \xsf$ responses, providing at least $h(\mathsf{N} - \mathsf{M}, \mathsf{M}/2) - \xsf$ linearly independent combinations. Meanwhile, the servers in $[\mathsf{M}]$ will have $\mathsf{M}/2 + \xsf$ responses, contributing at least $\mathsf{m} +\xsf$ linearly independent combinations. In total, these responses provide $h(\mathsf{N} - \mathsf{M}, \mathsf{M}/2)$ linearly independent combinations, which are sufficient to recover $F_1$ to $F_{2\mathsf{m}}$.

The number of linearly independent combinations transmitted by servers in $[\mathsf{M}+1:\mathsf{N}]$ is $h\left(\mathsf{N}-\mathsf{M}, \frac{\mathsf{M}}{2}\right)$, which spans the space containing $F_{\mathsf{m}+1}$ to $F_{2\mathsf{m}}$. Additionally, servers in $[\mathsf{M}]$ transmit $2\mathsf{m}$ linearly independent combinations, also spanning the space of $F_{\mathsf{m}+1}$ to $F_{2\mathsf{m}}$. Therefore, the total number of transmitted linearly independent combinations is $h(\mathsf{N},\mathsf{M})= 2\mathsf{m} + h\left(\mathsf{N}-\mathsf{M}, \frac{\mathsf{M}}{2}\right) - \mathsf{m} = h\left(\mathsf{N}-\mathsf{M}, \frac{\mathsf{M}}{2}\right) + \mathsf{m}$, which matches~\eqref{eq:M is even}.
\\

\subsection{general scheme 3}
\label{general scheme 3}
We now consider the $(\mathsf{N}, \mathsf{M})$ non-secure problem, where $1.5 \mathsf{M} \leq \mathsf{N} < 2\mathsf{M}$, $\mathsf{M}$ is odd, and $\mathsf{M} \geq 2\mathsf{m} + 1$. We aim to construct a scheme (Scheme~3) to prove~\eqref{eq:M is odd}.
We also define that $\mathsf{N}=2\mathsf{M}-\sf y$. 

{\it Data assignment phase.}
The assignment is shown at the top of the next page.
\begin{figure*}
\begin{align*}
\begin{array}{rl|c|c|c|c|c|c|c|c|c|c|c|c|}\cline{3-3}\cline{4-4}\cline{5-5}\cline{6-6}\cline{7-7}\cline{8-8}\cline{9-9}\cline{10-10}\cline{11-11}\cline{12-12}\cline{13-13}\cline{14-14}
&&\rule{0pt}{1.2em}\mbox{\small server} &\rule{0pt}{1.2em}\mbox{\small \negmedspace\negmedspace 1 \negmedspace\negmedspace} & \mbox{\small \negmedspace\negmedspace $\cdots$ \negmedspace\negmedspace} &  \rule{0pt}{1.2em}\mbox{\small \negmedspace\negmedspace $\sf y$ \negmedspace\negmedspace}& \rule{0pt}{1.2em}\mbox{\small \negmedspace\negmedspace  $\sf y +1$ \negmedspace\negmedspace}&  \rule{0pt}{1.2em}\mbox{\small \negmedspace\negmedspace  $\sf y+2$ \negmedspace\negmedspace}&  \rule{0pt}{1.2em}\mbox{\small \negmedspace\negmedspace $\cdots$ \negmedspace\negmedspace} &  \rule{0pt}{1.2em}\mbox{\small \negmedspace\negmedspace $\mathsf{M}$ \negmedspace\negmedspace} &\rule{0pt}{1.2em}\mbox{\small \negmedspace\negmedspace  $\mathsf{M}$+1 \negmedspace\negmedspace} & \rule{0pt}{1.2em}\mbox{\small \negmedspace\negmedspace  $\mathsf{M}+2$ \negmedspace\negmedspace} & \rule{0pt}{1.2em}\mbox{\small \negmedspace\negmedspace $\cdots$\negmedspace\negmedspace }& \rule{0pt}{1.2em}\mbox{\small \negmedspace\negmedspace $\mathsf{N}$\negmedspace\negmedspace} \\ 
\cline{3-3}\cline{4-4}\cline{5-5}\cline{6-6}\cline{7-7}\cline{8-8}\cline{9-9}\cline{10-10}\cline{11-11}\cline{12-12}\cline{13-13}\cline{14-14}
&& &D_1& \cdots       &   D_1 &  D_{1} &  D_{1} & \cdots &  D_{1}   &  D_{\frac{\mathsf{M}-1}{2}+1}   &  D_{\frac{\mathsf{M}-1}{2}+1} &  \cdots  &  D_{\frac{\mathsf{M}-1}{2}+1}  \\
&& &\cdots & \cdots & \cdots &  \cdots & \cdots  & \cdots   & \cdots  & \cdots & \cdots & \cdots & \cdots  \\ 
&& \mbox{\small dataset}&D_{\frac{\mathsf{M}-1}{2}} &    \cdots    & D_{\frac{\mathsf{M}-1}{2}}  & D_{\frac{\mathsf{M}-1}{2}} & D_{\frac{\mathsf{M}-1}{2}}    & \cdots  & D_{\frac{\mathsf{M}-1}{2}} &   D_{\mathsf{M}}  &   D_{\mathsf{M}}  &  \cdots  &  D_{\mathsf{M}}  \\
&& &D_{\frac{\mathsf{M}-1}{2}+1}&  \cdots  & D_{\frac{\mathsf{M}-1}{2}+1}  & D_{\mathsf{M}+1} & D_{\mathsf{M}+2} & \cdots & D_{\mathsf{N}}& D_{\mathsf{M}+1}  &  D_{\mathsf{M}+2} &   \cdots  &  D_{\mathsf{N}} \\ 
&& &\cdots &  \cdots       & \cdots & \cdots & \cdots   & \cdots & \cdots & \cdots & \cdots   & \cdots & \cdots \\
&& &D_{\mathsf{M}} &   \cdots  &  D_{\mathsf{M}} & D_{\frac{3\mathsf{M}+1}{2}} & D_{\frac{3\mathsf{M}+1}{2} +1}  & \cdots & D_{\frac{\mathsf{M}+1}{2}-1} &  D_{\frac{3\mathsf{M}-1}{2}}  &  D_{\frac{3\mathsf{M}-1}{2}+1}   &  \cdots  &  D_{\frac{\mathsf{M}-1}{2}-1}\\  
\cline{3-3}\cline{4-4}\cline{5-5}\cline{6-6}\cline{7-7}\cline{8-8}\cline{9-9}\cline{10-10}\cline{11-11}\cline{12-12}\cline{13-13}\cline{14-14}
\end{array} 
\end{align*}
\end{figure*}
In this assignment, we split the $\mathsf{N}$ datasets into three sections. The first section includes $D_1, \ldots, D_t$ (The reason for choosing $t = \frac{\mathsf{M} - 1}{2}$ is provided in \cite{wan2022secure}.), and these datasets are assigned to servers in $[\mathsf{M}]$. The second section, containing $D_{t+1}, \ldots, D_{\mathsf{M}}$, is assigned to servers in $[\sf y] \cup [\mathsf{M}+1 : \mathsf{N}]$. The third section, consisting of $D_{\mathsf{M}+1}, \ldots, D_{\mathsf{N}}$, is cyclically allocated to servers in $[\sf y+1 : \mathsf{M}]$, where each server receives $\mathsf{M} - t$ adjacent datasets within $\left[{\frac{\mathsf{M} - 1}{2} + 1} : \mathsf{N} \right]$. Additionally, datasets $D_{\mathsf{M}+1}, \ldots, D_{\mathsf{N}}$ are assigned to servers in $[\mathsf{M}+1 : \mathsf{N}]$ in a cyclic manner such that each server receives $t$ consecutive datasets within $\left[{\frac{\mathsf{M} - 1}{2} + 1} : \mathsf{N} \right]$.

  {\it Computing phase.}
  We first divide each message $W_k$ for $k \in [\Ksf]$ into $\mathsf{m}$ equal-length, non-overlapping sub-messages, denoted as $W_k = \{W_{k,j} : j \in [\mathsf{m}]\}$.
We design the computing phase so that the total number of linearly independent combinations computed by all servers is $\ \frac {\mathsf{M} + 1}{2} - \sf y + 2\mathsf{m}$. We let each server send a linear combination of messages so that the user can recover ${\bf F}' \ [W_{1,1}; \ldots; W_{\mathsf{N},\mathsf{m}}]$ from the responses of any $N_{\rm r} = \mathsf{N} - \mathsf{M} +\mathsf{m}$ servers, where \( {\bf F}' \) is shown at the top of the next page in \eqref{eq:general form of F'}.
\begin{figure*}
\begin{equation}
\scalebox{0.8}{$
{\bf F}' = \begin{bmatrix}
 \mathbf{f}_1 \\
 \vdots \\
 \mathbf{f}_{\frac{\mathsf{M}+1}{2}-\sf y+2\mathsf{m}} 
 \end{bmatrix}
 =
\left[
\begin{array}{c:c:c:c}
 ({\bf F})_{2 \times \mathsf{N}}  & {\bf 0}_{2 \times \mathsf{N}}  & \cdots & {\bf 0}_{2 \times \mathsf{N}}   \\ \hdashline
{\bf 0}_{2 \times \mathsf{N}} &  ({\bf F})_{2 \times \mathsf{N}}   & \cdots & {\bf 0}_{2 \times \mathsf{N}}   \\ \hdashline 
 \vdots   & \vdots  &  \ddots & \vdots \\ \hdashline
 {\bf 0}_{2 \times \mathsf{N}} &   {\bf 0}_{2 \times \mathsf{N}}    & \cdots &  ({\bf F})_{2 \times \mathsf{N}} \\ \hdashline
 ({\bf V}_{1})_{ (\frac{\mathsf{M} + 1}{2} - \sf y) \times \mathsf{N}}  &  ({\bf V}_{2})_{ (\frac{\mathsf{M} + 1}{2} - \sf y) \times \mathsf{N}}  &   \cdots   &   ({\bf V}_{\mathsf{m}})_{ (\frac{\mathsf{M} + 1}{2} - \sf y) \times \mathsf{N}} 
 \end{array}
\right]_{(\frac{\mathsf{M} + 1}{2} - \sf y + 2\mathsf{m}) \times \mathsf{N}\mathsf{m}}.
$}
\label{eq:general form of F'}
\end{equation}

\end{figure*}

We design the demand matrix 
\begin{equation}
\bf F = 
\begin{bmatrix}
\tikzmark{left4F} \textcolor{white}{0} 1 , \ldots , 1 \textcolor{white}{0} \tikzmark{right4F} & 
\tikzmark{left5F} \textcolor{white}{0} 1 , \ldots , 1 \textcolor{white}{0} \tikzmark{right5F} & 
\tikzmark{left6F} \textcolor{white}{0} 1 , \ldots , 1 \textcolor{white}{0} \tikzmark{right6F} \\
\textcolor{white}{0} 0 , \ldots , 0 \textcolor{white}{0} & 
\textcolor{white}{0} 2 , \ldots , 2 \textcolor{white}{0} & 
\textcolor{white}{0} 1 , \ldots , 1 \textcolor{white}{0} 
\end{bmatrix}
\label{eq: general of F}
\end{equation}

\DrawboxF[thick, black, dashed]{left4F}{right4F}{\textcolor{black}{\footnotesize${\bf C}_1$}}
\DrawboxF[thick, red, dashed]{left5F}{right5F}{\textcolor{red}{\footnotesize${\bf C}_2$}}
\DrawboxF[thick, blue, dashed]{left6F}{right6F}{\textcolor{blue}{\footnotesize${\bf C}_3$}}\\

We divide matrix ${\bf F}$ into three column-wise sub-matrices: ${\bf C}_1$, containing $\frac{\mathsf{M}-1}{2}$ columns, corresponding to the messages in $\left[\frac{\mathsf{M}-1}{2} \right]$, ${\bf C}_2$, containing $\frac{\mathsf{M}+1}{2}$ columns, corresponding to the messages in $\left[\frac{\mathsf{M}+1}{2} : \mathsf{M} \right]$, and ${\bf C}_3$, containing $(\mathsf{N} - \mathsf{M})$ columns, corresponding to the messages in $[\mathsf{M}+1 : \mathsf{N}]$.

Let the matrix ${\bf V}_i$, for $i \in [\mathsf{m}]$, be defined as
\begin{equation}
{\bf V}_i = 
\begin{bmatrix}
\tikzmark{left4V} \textcolor{white}{0} 0 , \ldots , 0 \textcolor{white}{0} \tikzmark{right4V} & 
\tikzmark{left5V} \textcolor{white}{0} 0 , \ldots , 0 \textcolor{white}{0} \tikzmark{right5V} & 
\tikzmark{left6V} \textcolor{white}{0} * , \ldots , * \textcolor{white}{0} \tikzmark{right6V} \\
\vdots & \vdots & \vdots \\
\textcolor{white}{0} 0 , \ldots , 0 \textcolor{white}{0} & 
\textcolor{white}{0} 0 , \ldots , 0 \textcolor{white}{0} & 
\textcolor{white}{0} * , \ldots , * \textcolor{white}{0} 
\end{bmatrix}
\label{eq:general form of V_i}
\end{equation}

\DrawboxV[thick, black, dashed]{left4V}{right4V}{\textcolor{black}{\footnotesize${\bf C}_1$}}
\DrawboxV[thick, red, dashed]{left5V}{right5V}{\textcolor{red}{\footnotesize${\bf C}_2$}}
\DrawboxV[thick, blue, dashed]{left6V}{right6V}{\textcolor{blue}{\footnotesize${\bf C}_3$}}

In the matrix ${\bf V}_i$, the entries denoted by $*$ are the unknowns we need to design. And ${\bf C}_1$, ${\bf C}_2$, and ${\bf C}_3$ have column counts consistent with the definitions above.

Now, we focus on the servers in $[\sf y]$, which have the same datasets and do not contain datasets from $[\mathsf{M}+1 : \mathsf{N}]$. From these servers, we aim to recover:
\begin{align*}
    F_1 - F_2 &= W_{1,1} + \cdots + W_{t,1} - W_{t+1,1} - \cdots - W_{\mathsf{N},1}, \\
    &\vdots \\
    F_{2\mathsf{m} - 1} - F_{2\mathsf{m}} &= W_{1,\mathsf{m}} + \cdots + W_{t,\mathsf{m}} - W_{t+1,\mathsf{m}} - \cdots - W_{\mathsf{N},\mathsf{m}}.
\end{align*}
Thus, we let the servers in $[\sf y]$ each compute a random linear combination of the $m$ linearly independent combinations $(F_1 - F_2), \ldots, (F_{2\mathsf{m} - 1} - F_{2\mathsf{m}})$.

For the servers in $[M+1 : N]$ that do not contain datasets from $[t]$, we design their transmissions so that the user can recover $F_2, F_4, \ldots, F_{2\mathsf{m}}, F_{2\mathsf{m}+1}, \ldots, F_{\frac{\mathsf{M}+1}{2} - \sf y + 2\mathsf{m}}$ from any $\frac{\mathsf{M}+1}{2} - \sf y + \mathsf{m}$ responses.
 For $n \in [\mathsf{M}+1, \mathsf{N}]$, we let server $n$ compute
\begin{align}
{\bf s}_n  & [ F_2; F_4; \ldots; F_{2\mathsf{m}}; F_{2\mathsf{m}+1}; \ldots; F_{\frac{\mathsf{M}+1}{2} - \sf y + 2\mathsf{m}} ]  \nonumber \\
& \text{× }  [ W_{1,1}; W_{2,1}; \ldots; W_{\mathsf{N},\mathsf{m}} ] .
\label{eq:transmission of second class}
\end{align}
We design ${\bf s}_{n} $ as follows.
Notice that $W_1, \ldots, W_{\frac{\mathsf{M}-1}{2}}$ can be computed by server $n$; and that in the linear combination~\eqref{eq:transmission of second class}, the coefficients of $W_{\frac{\mathsf{M}-1}{2}+1}, \ldots, W_{\mathsf{M}}$ are $0$. 
Hence, to guarantee that in~\eqref{eq:transmission of second class} the coefficients of the messages which server $n$ cannot compute are $0$, we only need to consider the messages in $W_{\mathsf{M}+1}, \ldots, W_{\mathsf{N}}$, corresponding to the columns in ${\bf C}_3$ in the matrix, with a total of $\mathsf{m} (\mathsf{N} - \mathsf{M})$ columns.
By extracting the corresponding rows and columns from the matrix \( {\bf F}' \), we obtain \( {\bf F}_{\bf s} \), as shown at the top of the next page in \eqref{eq:general form of F_s},
\begin{figure*}
\begin{equation}
\scalebox{0.8}{$
 {\bf F}_{\bf s} = 
\left[
\begin{array}{c:c:c:c}
{\bf 1}_{1 \times (\mathsf{N}-\mathsf{M})}  & {\bf 0}_{1 \times (\mathsf{N}-\mathsf{M})}  & \cdots & {\bf 0}_{1 \times (\mathsf{N}-\mathsf{M})}   \\ \hdashline
{\bf 0}_{1 \times (\mathsf{N}-\mathsf{M})} &  {\bf 1}_{1 \times (\mathsf{N}-\mathsf{M})}   & \cdots & {\bf 0}_{1 \times (\mathsf{N}-\mathsf{M})}   \\ \hdashline 
 \vdots   & \vdots  &  \ddots & \vdots \\ \hdashline
 {\bf 0}_{1 \times (\mathsf{N}-\mathsf{M})} &   {\bf 0}_{1 \times (\mathsf{N}-\mathsf{M})}    & \cdots &  {\bf 1}_{1 \times (\mathsf{N}-\mathsf{M})} \\ \hdashline
 ({\bf V}'_{1})_{ (\frac{\mathsf{M} + 1}{2} - \sf y) \times (\mathsf{N}-\mathsf{M})}  &  ({\bf V}'_{2})_{ (\frac{\mathsf{M} + 1}{2} - \sf y) \times (\mathsf{N}-\mathsf{M})}  &   \cdots   &   ({\bf V}'_{\mathsf{m}})_{ (\frac{\mathsf{M} + 1}{2} - \sf y) \times (\mathsf{N}-\mathsf{M})} 
 \end{array}
\right]_{(\frac{\mathsf{M} + 1}{2} - \sf y + \mathsf{m}) \times \mathsf{m}(\mathsf{N}-\mathsf{M})}
$}
\label{eq:general form of F_s}
\end{equation}

\end{figure*}
where ${\bf V}'_i$ is the sub-matrix formed by the last $\mathsf{N} - \mathsf{M}$ columns of ${\bf V}_i$.
Notice that the datasets in $[\mathsf{M}+1 : \mathsf{N}]$ are placed in a cyclic assignment among the servers in $[\mathsf{M}+1 : \mathsf{N}]$. Each server cannot compute $\mathsf{N} - \mathsf{M} - \frac{\mathsf{M} - 1}{2}$ of the datasets in $[\mathsf{M}+1 : \mathsf{N}]$. By extracting the corresponding columns from the matrix ${\bf F}_{\bf s}$, we form a submatrix ${\bf F}_{{\bf s}_n}$, with dimensions 
$\left(\frac{\mathsf{M} + 1}{2} - \sf y + \mathsf{m}\right) \times \mathsf{m} \left(\mathsf{N} - \frac{3\mathsf{M} - 1}{2}\right)$.
Our desired vector ${\bf s}_n$ is then the left null vector of this matrix.
If the number of rows in the matrix ${\bf F}_{{\bf s}_n}$ is greater than the number of columns, we can directly assign random values to the entries denoted by $*$ as in \cite{wan2022secure} to obtain ${\bf s}_n$. If the number of rows is less than or equal to the number of columns, we apply the interference alignment strategy from \cite{huang2023ISITversion} to design the values of the $*$ entries in order to obtain ${\bf s}_n$.

Finally, we focus on the servers in $[\sf y+1 : \mathsf{M}]$. We design their transmissions so that the user can recover $2F_1 - F_2, 2F_3 - F_4, \ldots, 2F_{2\mathsf{m}-1} - F_{2\mathsf{m}}, F_{2\mathsf{m}+1}, \ldots, F_{\frac{\mathsf{M}+1}{2} - \sf y + 2\mathsf{m}}$ from any $\frac{\mathsf{M}+1}{2} - \sf y + \mathsf{m}$ responses.
 For $n \in [\sf y+1, \mathsf{M}]$, we also let server $n$ compute
\begin{align}
{\bf s}_n & [ 2F_1 - F_2; \ldots; 2F_{2\mathsf{m}-1} - F_{2\mathsf{m}}; F_{2\mathsf{m}+1}; \ldots; F_{\frac{\mathsf{M}+1}{2} - \sf y + 2\mathsf{m}} ]  \nonumber \\
& \text{× }[ W_{1,1}; W_{2,1}; \ldots; W_{\mathsf{N},\mathsf{m}} ].
\label{eq:transmission of first class}
\end{align}
     We design ${\bf s}_{n} $ as follows.
Notice that $W_1,\ldots,W_{\frac{\mathsf{M}-1}{2}}$ can be computed by server $n$; and that in the linear combination~\eqref{eq:transmission of first class} the coefficients of   $W_{\frac{\mathsf{M}-1}{2}+1}, \ldots, W_{\mathsf{M}}$ are $0$. 
Hence, in order to guarantee that in~\eqref{eq:transmission of first class} the coefficients of the messages which server $n$ cannot compute are $0$, we only need to consider the messages in $W_{\mathsf{M}+1}, \ldots, W_{\mathsf{N}}$,corresponding to the columns in ${\bf C}_3$ in the matrix, with a total of $\mathsf{m} (\mathsf{N} - \mathsf{M})$ columns.By extracting the corresponding rows and columns from the matrix \( {\bf F}' \), we obtain \( {\bf F}'_{\bf s} \), as shown at the top of the next page in \eqref{eq:general form of F'_s}.
\begin{figure*}
\begin{equation}
\scalebox{0.8}{$
 {\bf F}'_{\bf s} = 
\left[
\begin{array}{c:c:c:c}
{\bf 1}_{1 \times (\mathsf{N}-\mathsf{M})}  & {\bf 0}_{1 \times (\mathsf{N}-\mathsf{M})}  & \cdots & {\bf 0}_{1 \times (\mathsf{N}-\mathsf{M})}   \\ \hdashline
{\bf 0}_{1 \times (\mathsf{N}-\mathsf{M})} &  {\bf 1}_{1 \times (\mathsf{N}-\mathsf{M})}   & \cdots & {\bf 0}_{1 \times (\mathsf{N}-\mathsf{M})}   \\ \hdashline 
 \vdots   & \vdots  &  \ddots & \vdots \\ \hdashline
 {\bf 0}_{1 \times (\mathsf{N}-\mathsf{M})} &   {\bf 0}_{1 \times (\mathsf{N}-\mathsf{M})}    & \cdots &  {\bf 1}_{1 \times (\mathsf{N}-\mathsf{M})} \\ \hdashline
 ({\bf V}'_{1})_{ (\frac{\mathsf{M} + 1}{2} - \sf y) \times (\mathsf{N}-\mathsf{M})}  &  ({\bf V}'_{2})_{ (\frac{\mathsf{M} + 1}{2} - \sf y) \times (\mathsf{N}-\mathsf{M})}  &   \cdots   &   ({\bf V}'_{\mathsf{m}})_{ (\frac{\mathsf{M} + 1}{2} - \sf y) \times (\mathsf{N}-\mathsf{M})} 
 \end{array}
\right]_{(\frac{\mathsf{M} + 1}{2} - \sf y + \mathsf{m}) \times \mathsf{m}(\mathsf{N}-\mathsf{M})}.
$}
\label{eq:general form of F'_s}
\end{equation}

\end{figure*}
It is easy to observe that ${\bf F}'_{\bf s} = {\bf F}_{\bf s}$.Notice that the datasets in $[\mathsf{M}+1, \mathsf{N}]$ are also cyclically assigned to the servers in $[\sf y+1 : \mathsf{M}]$. For $n \in [\sf y+1 : \mathsf{M}]$, server $n$ cannot compute ${\mathsf{N} - \frac{3\mathsf{M} + 1}{2}}$ datasets from $[\mathsf{M}+1, \mathsf{N}]$, and these datasets are also not computable by server $n + \mathsf{M} - \sf y$. Since ${\bf F}'_{\bf s} = {\bf F}_{\bf s}$, we can  set ${\bf s}_n = {\bf s}_{n + \mathsf{M} - \sf y}$.

  {\it Decoding phase.}
The user receives responses from $N_{\rm r} = \mathsf{N} - \mathsf{M} + \mathsf{m}$ servers, of which $x$ responses are from servers in $[\sf y]$. Depending on the value of $\xsf$, we can categorize the scenarios into the following two cases:
\begin{itemize}
\item {\it Case 1: $\xsf \leq \mathsf{m}$.} In this case, the user receives $\xsf$ linearly independent combinations sent by the servers in $[\sf y]$. The user still needs to collect an additional $2\mathsf{m} - \xsf + \mathsf{N} - \frac{3\mathsf{M} - 1}{2}$ linearly independent combinations from any $(\mathsf{N} - \mathsf{M} - \xsf + \mathsf{m})$ servers in $[\sf y + 1 : \mathsf{N}]$.
In the worst-case scenario, the servers in one of the groups, either $[\sf y + 1 : \mathsf{M}]$ or $[\mathsf{M} + 1 : \mathsf{N}]$, all respond, transmitting a total of $\mathsf{m} + \mathsf{N} - \frac{3\mathsf{M} - 1}{2}$ linearly independent combinations. Additionally, $\mathsf{m} - \xsf$ servers from the other group respond, providing another $\mathsf{m} - \xsf$ linearly independent combinations. 

Together, this results in a total of $2\mathsf{m} + \mathsf{N} - \frac{3\mathsf{M} - 1}{2}$ linearly independent combinations, sufficient to fully recover the demand matrix ${\bf F}'$.

\item {\it Case 2: $\mathsf{m} < \xsf \leq \sf y$.} In this case, the user receives $\mathsf{m}$ linearly independent combinations sent by the servers in $[\sf y]$. We consider the worst-case scenario, where $x = \sf y$, meaning that the user still needs to collect an additional $\mathsf{m} + \mathsf{N} - \frac{3\mathsf{M} - 1}{2}$ linearly independent combinations from any $2\mathsf{N} - 3\mathsf{M} + \mathsf{m}$ servers in $[\sf y + 1 : \mathsf{N}]$.
Since $\mathsf{N} \geq \frac{3\mathsf{M} + 1}{2}$, it follows that $2\mathsf{N} - 3\mathsf{M} + \mathsf{m} \geq \mathsf{m} + \mathsf{N} - \frac{3\mathsf{M} - 1}{2}$. Additionally, any $\mathsf{m} + \mathsf{N} - \frac{3\mathsf{M} - 1}{2}$ servers in either of the groups $[\sf y + 1 : \mathsf{M}]$ or $[\mathsf{M} + 1 : \mathsf{N}]$ compute linearly independent combinations. Therefore, it is possible to receive $\mathsf{m} + \mathsf{N} - \frac{3\mathsf{M} - 1}{2}$ linearly independent combinations from any $2\mathsf{N} - 3\mathsf{M} + \mathsf{m}$ servers in $[\sf y + 1 : \mathsf{N}]$.
\end{itemize}
Thus, the user obtains a total of $2\mathsf{m} + \mathsf{N} - \frac{3\mathsf{M} - 1}{2}$ linearly independent combinations, which suffice to reconstruct the demand matrix ${\bf F}'$.

 By the above scheme,  the number of linearly independent transmissions by all servers is equal to the number of rows in ${\bf F}'$, i.e.,
$\frac{\mathsf{M}+1}{2}-\sf y+2\mathsf{m} =\mathsf{N}- \frac{3\mathsf{M}-1}{2}+2\mathsf{m} $, coinciding with~\eqref{eq:M is odd}.

\subsection{general scheme 4}
\label{general scheme 4}
In this section, we examine the case where $\mathsf{M} < \mathsf{N} < 1.5 \mathsf{M}$, with $\mathsf{M} \geq 2\mathsf{m}$, and construct a recursive scheme (Scheme~4) that aims to demonstrate~\eqref{eq:less than 1.5M}. This recursive scheme builds on the solution for the $(\mathsf{M}, 2\mathsf{M}-\mathsf{N})$ non-secure problem, assuming that this solution has been previously developed, and provides a total of $h(\mathsf{M}, 2\mathsf{M}-\mathsf{N})$ linearly independent combinations of messages.

{\it Data assignment phase.} We begin by allocating datasets $D_{1}, \ldots, D_{\mathsf{N}-\mathsf{M}}$ to each server in $[\mathsf{M}]$. Subsequently, datasets $D_{\mathsf{N}-\mathsf{M}+1}, \ldots, D_{\mathsf{N}}$ are assigned to each server in $[\mathsf{M}+1:\mathsf{N}]$, ensuring that each server in $[\mathsf{M}+1:\mathsf{N}]$ has access to $\mathsf{M}$ datasets, while each server in $[\mathsf{M}]$ holds fewer than $\mathsf{M}$ datasets, specifically $\mathsf{N} - \mathsf{M}$. Additionally, each dataset in $[\mathsf{N}-\mathsf{M}+1 : \mathsf{N}]$ is assigned to $\mathsf{N} - \mathsf{M} < \mathsf{M}$ servers. Therefore, in the next step, we assign each dataset $D_k$ for $k \in [\mathsf{N} - \mathsf{M} + 1 : \mathsf{N}]$ to $2\mathsf{M} - \mathsf{N}$ servers in $[\mathsf{M}]$, such that each server in $[\mathsf{M}]$ obtains $2\mathsf{M} - \mathsf{N}$ datasets in $[\mathsf{N} - \mathsf{M} + 1 : \mathsf{N}]$. This allocation follows the assignment phase of the proposed scheme for the $(\mathsf{M}, 2\mathsf{M}-\mathsf{N})$ non-secure problem.

{\it Computing phase.} Each message $W_k$ for $k \in [\Ksf]$ is divided into $\mathsf{m}$ equal-length, non-overlapping sub-messages, noted as $W_k = \{W_{k,j} : j \in [\mathsf{m}]\}$. First, we focus on the $(\mathsf{M}, 2\mathsf{M}-\mathsf{N})$ non-secure problem, where the messages are defined as $W^{\prime}_1, \ldots, W^{\prime}_{\mathsf{M}}$. Under this scheme, each server computes a linear combination of these $\mathsf{M}$ messages. Letting the total number of linearly independent combinations across all servers be $h(\mathsf{M}, 2\mathsf{M}-\mathsf{N})$, we can represent these $h(\mathsf{M}, 2\mathsf{M}-\mathsf{N})$ combinations as  
\begin{align}
{\bf F}' \  [W^{\prime}_{1,1}; \ldots; W^{\prime}_{\mathsf{M},\mathsf{m}}].\label{eq:F_1}
\end{align}
Each transmission from a server $n^{\prime} \in \left[ \mathsf{M} \right]$ is represented as
$$
\mathbf{s}_{n^{\prime}} \ {\bf F}'  \  [W^{\prime}_{1,1}; \ldots; W^{\prime}_{\mathsf{M},\mathsf{m}}].
$$

Returning to the \( (\mathsf{N}, \mathsf{M}) \) non-secure problem, we set up the server transmissions so that, together, they produce \( h(\mathsf{M}, 2\mathsf{M} - \mathsf{N}) \) linearly independent combinations, represented as \( {\bf F} [W_{1,1}; \ldots; W_{\mathsf{N}, \mathsf{m}}] \), where \( {\bf F} \) is shown at the top of the next page in \eqref{eq:general form of F}.
\begin{figure*}[ht]
\begin{equation}
\resizebox{\textwidth}{!}{$ 
{\bf F} = 
\left[
\begin{array}{c:c:c:c:c:c:c}
{\bf 1}_{1 \times (\mathsf{N} - \mathsf{M})} & {\bf 1}_{1 \times \mathsf{M}} & {\bf 0}_{1 \times (\mathsf{N} - \mathsf{M})} & {\bf 0}_{1 \times \mathsf{M}} & \cdots & {\bf 0}_{1 \times (\mathsf{N} - \mathsf{M})} & {\bf 0}_{1 \times \mathsf{M}} \\ \hdashline
{\bf 0}_{1 \times (\mathsf{N} - \mathsf{M})} & {\bf 0}_{1 \times \mathsf{M}} & {\bf 1}_{1 \times (\mathsf{N} - \mathsf{M})} & {\bf 1}_{1 \times \mathsf{M}} & \cdots & {\bf 0}_{1 \times (\mathsf{N} - \mathsf{M})} & {\bf 0}_{1 \times \mathsf{M}} \\ \hdashline 
\vdots & \vdots & \vdots & \vdots & \ddots & \vdots & \vdots \\ \hdashline
{\bf 0}_{1 \times (\mathsf{N} - \mathsf{M})} & {\bf 0}_{1 \times \mathsf{M}} & {\bf 0}_{1 \times (\mathsf{N} - \mathsf{M})} & {\bf 0}_{1 \times \mathsf{M}} & \cdots & {\bf 1}_{1 \times (\mathsf{N} - \mathsf{M})} & {\bf 1}_{1 \times \mathsf{M}} \\ \hdashline
({\bf A})_{(h(\mathsf{M}, 2\mathsf{M}-\mathsf{N}) - \mathsf{m}) \times (\mathsf{N} - \mathsf{M})} & ({\bf V}_{1})_{(h(\mathsf{M}, 2\mathsf{M}-\mathsf{N}) - \mathsf{m}) \times \mathsf{M}} & ({\bf A})_{(h(\mathsf{M}, 2\mathsf{M}-\mathsf{N}) - \mathsf{m}) \times (\mathsf{N} - \mathsf{M})} & ({\bf V}_{2})_{(h(\mathsf{M}, 2\mathsf{M}-\mathsf{N}) - \mathsf{m}) \times \mathsf{M}} & \cdots & ({\bf A})_{(h(\mathsf{M}, 2\mathsf{M}-\mathsf{N}) - \mathsf{m}) \times (\mathsf{N} - \mathsf{M})} & ({\bf V}_{\mathsf{m}})_{(h(\mathsf{M}, 2\mathsf{M}-\mathsf{N}) - \mathsf{m}) \times \mathsf{M}}
\end{array}
\right]_{h(\mathsf{M}, 2\mathsf{M} - \mathsf{N}) \times \mathsf{m} \mathsf{N}}.$}
\label{eq:general form of F}
\end{equation}

\end{figure*}
Each row in the submatrix ${\bf A}$ contains identical random numbers. For each $i \in [\mathsf{m}]$, ${\bf V}_i$ shares elements with the corresponding elements in matrix ${\bf F}$, allowing matrix ${\bf F}$ to be viewed as a submatrix extracted from ${\bf F}'$ based on the columns in each ${\bf V}_i$.

For each server $n \in [\mathsf{M}]$, by the construction in the assignment phase, it has access to datasets $D_1, \ldots, D_{\mathsf{N} - \mathsf{M}}$, while the assignment for datasets $D_{\mathsf{N} - \mathsf{M} + 1}, \ldots, D_{\mathsf{N}}$ is based on the assignment for the $(\mathsf{M}, 2\mathsf{M} - \mathsf{N})$ non-secure problem. Thus, server $n$ computes
$ 
\mathbf{s}_{n} \ {\bf F} \ [W_1; \ldots; W_{\mathsf{N}}],
$
where $\mathbf{s}_n$ corresponds to the transmission vector of server $n$ in the $(\mathsf{M}, 2\mathsf{M} - \mathsf{N})$ problem.

For each server $n \in [\mathsf{M}+1 : \mathsf{N}]$, it cannot compute $W_1, \ldots, W_{\mathsf{N} - \mathsf{M}}$, which corresponds to the columns of submatrix ${\bf A}$ in ${\bf F}$. Extracting these columns creates a matrix ${\bf F}''$ of size $h(\mathsf{M}, 2\mathsf{M} - \mathsf{N}) \times \mathsf{m} (\mathsf{N} - \mathsf{M})$, whose rank is $\mathsf{m}$. Consequently, the left null space of ${\bf F}''$ includes $h(\mathsf{M}, 2\mathsf{M} - \mathsf{N}) - \mathsf{m}$ independent vectors. We let the transmission vector $\mathbf{s}_n$ for each server $n$ be a random linear combination of these vectors, using uniformly i.i.d. coefficients from $\mathbb{F}_{\qsf}$, so that each server $n$ computes $\mathbf{s}_n {\bf F} [W_1; \ldots; W_{\mathsf{N}}]$.

{\it Decoding phase.} Suppose $\Ac$ denotes the set of responding servers with $|\Ac| = \mathsf{N}_\mathrm{r}$, $\Ac_1 = \Ac \cap [\mathsf{M}]$, and $\Ac_2 = \Ac \setminus [\mathsf{M}]$, where $|\Ac_2| \leq \mathsf{N} - \mathsf{M} = \mathsf{N}_\mathrm{r} - \mathsf{m}$. When $\mathsf{M} \geq \mathsf{N} - \mathsf{M} + \mathsf{m} = \mathsf{N}_\mathrm{r}$, if $|\Ac_2| = 0$, the user can recover the first $\mathsf{m}$ rows of ${\bf F}$, leveraging the decodability of the $(\mathsf{M}, 2\mathsf{M} - \mathsf{N})$ scheme.

When $0 < |\Ac_2| \leq \mathsf{N} - \mathsf{M}$ and $\Ac_1$ alone does not suffice, the user receives at least $\mathsf{m}$ independent responses from $\Ac_1$. Let the linearly independent combinations from $\Ac_1$ be denoted by $\lambda_1$. In addition to the responses from $\Ac_1$, the user will receive responses from at least $h(\mathsf{M}, 2\mathsf{M} - \mathsf{N}) - \lambda_1$ servers in $[\mathsf{M}+1 : \mathsf{N}]$. Since each server sends an independent combination of $h(\mathsf{M}, 2\mathsf{M} - \mathsf{N}) - \mathsf{m}$, any $h(\mathsf{M}, 2\mathsf{M} - \mathsf{N}) - \lambda_1$ responses in $[\mathsf{M}+1 : \mathsf{N}]$ are linearly independent with high probability. Combining with $\lambda_1$, this totals $h(\mathsf{M}, 2\mathsf{M} - \mathsf{N})$ combinations, allowing the user to recover ${\bf F} \begin{bmatrix} W_{1,1}; \ldots; W_{\mathsf{N}, \mathsf{m}} \end{bmatrix}$.

In summary, $h(\mathsf{N}, \mathsf{M}) = h(\mathsf{M}, 2\mathsf{M} - \mathsf{N})$, consistent with~\eqref{eq:less than 1.5M}.

\section{The applicable range of Interference Alignment in general scheme 3
}

Recall that in \eqref{eq:general form of F_s}, ${\bf F}_{\bf s}$ is a matrix of dimensions $\left( \frac{\mathsf{M} + 1}{2} - \sf y + \mathsf{m} \right) \times \mathsf{m}(\mathsf{N} - \mathsf{M})$. We use $\mathsf{N}-\mathsf{M}$ workers to compute this required matrix, where each worker can compute $\frac{\mathsf{M} - 1}{2}$ datasets. If we proceed with interference alignment individually for each worker as in Example \ref{ex:scheme 3 example}, we should generate $(\mathsf{m}-1)\left(\mathsf{N} - \frac{3\mathsf{M}}{2} - \frac{1}{2}\right)$ linearly independent linear equations for each worker. Thus, in total, we generate $(\mathsf{N}-\mathsf{M})(\mathsf{m}-1)\left(\mathsf{N} - \frac{3\mathsf{M}}{2} - \frac{1}{2}\right)$ linear equations. The resulting matrix ${\bf E}$ has dimensions $ \left[ (\mathsf{N}-\mathsf{M})(\mathsf{m}-1)\left(\mathsf{N} - \frac{3\mathsf{M}}{2} - \frac{1}{2}\right) \right] \times \mathsf{m}(\mathsf{N}-\mathsf{M})$. By construction, these $(\mathsf{N}-\mathsf{M})(\mathsf{m}-1)\left(\mathsf{N} - \frac{3\mathsf{M}}{2} - \frac{1}{2}\right)$ linear equations are linearly independent, and thus, ${\bf E}$ is of full rank. The left null space of ${\bf E}^{\text{\rm T}}$ contains only $\mathsf{m}(\mathsf{N}-\mathsf{M}) - (\mathsf{N}-\mathsf{M})(\mathsf{m}-1)\left(\mathsf{N} - \frac{3\mathsf{M}}{2} - \frac{1}{2}\right)$ linearly independent vectors. However, the number of rows of ${\bf F}_{\bf s}$ is $\left( \frac{\mathsf{M} + 1}{2} - \sf y + \mathsf{m} \right) = \mathsf{N} - \frac{3\mathsf{M}}{2} + \frac{1}{2} + \mathsf{m}$. Therefore, when 
$$ \mathsf{m}(\mathsf{N}-\mathsf{M}) - (\mathsf{N}-\mathsf{M})(\mathsf{m}-1)\left(\mathsf{N} - \frac{3\mathsf{M}}{2} - \frac{1}{2}\right) < \mathsf{N} - \frac{3\mathsf{M}}{2} + \frac{1}{2} + \mathsf{m}, $$
i.e.,
$$ \mathsf{m}(\mathsf{N} - \mathsf{M} - 1) \leq \left[ (\mathsf{N} - \mathsf{M})(\mathsf{m} - 1) + 1 \right] \left( \mathsf{N} - \frac{3}{2} \mathsf{M} - \frac{1}{2} \right), $$
we need to use the interference alignment computing scheme, which jointly aligns the interferences across different workers.




\end{document}